\documentclass[journal]{IEEEtran}
\usepackage{graphicx}
\usepackage{subfigure}
\usepackage{cite}
\usepackage{bm}
\usepackage{amsmath}   % From the American Mathematical Society
\usepackage{amsthm}
\usepackage{cases}
\usepackage{amsfonts,amssymb}
\usepackage{cuted}\stripsep-3pt
\usepackage{lipsum}

\usepackage{color}
\usepackage[normalem]{ulem}
\usepackage{float}
\usepackage{stfloats}

\ifCLASSINFOpdf
\else
\fi
\begin{document}
%\title{OAM Channel Measurement, Modeling, and Characterization at 5.8 and 28 GHz}
\title{A New Channel Model for OAM Wireless Communication at 5.8 and 28 GHz}

\author{\IEEEauthorblockN{Runyu Lyu, \textit{Student Member, IEEE}, Wenchi Cheng, \textit{Senior Member, IEEE}, Muyao Wang, \textit{Student Member, IEEE}, Fan Qin, \textit{Senior Member, IEEE}, and Tony Q.S. Quek, \textit{Fellow, IEEE}~}% <-this % stops an unwanted space

%\author{\IEEEauthorblockN{Runyu Lyu, Wenchi Cheng, and Muyao Wang}% <-this % stops an unwanted space
%\IEEEauthorblockA{State Key Laboratory of Integrated Services Networks, Xidian University, Xi'an, China}
%\vspace{-20pt}
\thanks{%Manuscript received March 31, 2023; revised August 15, 2023; accepted October 19, 2023. This work was supported in part by National Key R\&D Program of
%China under Grant 2021YFC3002102, in part by the Key R\&D Plan of
%Shaanxi Province under Grant 2022ZDLGY05-09, and in part by Key Area
%Research and Development Program of Guangdong Province under Grant
%2020B0101110003. (Corresponding author: Wenchi Cheng.)

%Runyu Lyu, Wenchi Cheng, Muyao Wang, and Fan Qin are with School of Telecommunications Engineering, Xidian University, Xi¡¯an, 710071, China (e-mails: rylv@stu.xidian.edu.cn; wccheng@xidian.edu.cn). Tony Q.S. Quek is with Information Systems Technology and Design, Singapore University of Technology and Design, Singapore, 487372, Singapore. The corresponding author is Wenchi Cheng.
%
Runyu Lyu, Wenchi Cheng, Muyao Wang, and Fan Qin are with Xidian University, Xi'an, 710071, China (e-mails: rylv@stu.xidian.edu.cn; wccheng@xidian.edu.cn).

Tony Q.S. Quek is with Information Systems Technology and Design, Singapore University of Technology and Design, Singapore, 487372, Singapore (e-mails: tonyquek@sutd.edu.sg).
}
}

%{Shell \MakeLowercase{\textit{et al.}}: Bare Demo of IEEEtran.cls for IEEE Transactions on Magnetics Journals}



% for Transactions on Magnetics papers, we must declare the abstract and
% index terms PRIOR to the title within the \IEEEtitleabstractindextext
% IEEEtran command as these need to go into the title area created by
\IEEEtitleabstractindextext{%
%\vspace{-10pt}
\begin{abstract}
Orbital angular momentum (OAM) in electromagnetic (EM) waves can significantly enhance spectrum efficiency in wireless communications without requiring additional power, time, or frequency resources. Different OAM modes in EM waves create orthogonal channels, thereby improving spectrum efficiency. Additionally, OAM waves can more easily maintain orthogonality in line-of-sight (LOS) transmissions, offering an advantage over multiple-input and multiple-output (MIMO) technology in LOS scenarios. However, challenges such as divergence and crosstalk hinder OAM's efficiency. Additionally, channel modeling for OAM transmissions is still limited. A reliable channel model with balanced accuracy and complexity is essential for further system analysis. In this paper, we present a quasi-deterministic channel model for OAM channels in the 5.8 GHz and 28 GHz bands based on measurement data. Accurate measurement, especially at high frequencies like millimeter bands, requires synchronized RF channels to maintain phase coherence and purity, which is a major challenge for OAM channel measurement. To address this, we developed an 8-channel OAM generation device at 28 GHz to ensure beam integrity. By measuring and modeling OAM channels at 5.8 GHz and 28 GHz with a modified 3D geometric-based stochastic model (GBSM), this study provides insights into OAM channel characteristics, aiding simulation-based analysis and system optimization.
\end{abstract}

%\vspace{-5pt}
}

\maketitle

\IEEEdisplaynontitleabstractindextext

% \fi
\IEEEpeerreviewmaketitle

\vspace{15pt}
\section{Introduction}
Just as in rigid body motion where objects have angular momentum, electromagnetic (EM) waves also possess two types of angular momentum: spin angular momentum (SAM) and orbital angular momentum (OAM). In rigid body motion, SAM is the rotation around an object's own axis, like a spinning top, while OAM is a wobbling motion as the object spins, similar to the way a spinning top might tilt and wobble. For EM waves, SAM is the quantum spin and the rotation involving the polarization degrees of freedom of photons, whereas OAM is the rotation of the wave's field spatial distribution\cite{oam_light,oam_for_wireless_communication,oam_low_freq_radio}, meaning how the wave's shape or pattern rotates in space. SAM, achieved through polarization techniques, is already widely used in current wireless systems to increase the capacity and efficiency. Similarly, OAM can also be used to enhance spectrum efficiency. OAM-carrying radio frequency (RF) waves with different OAM modes are orthogonal to each other, allowing the creation of multiple independent channels. This orthogonality enables the enhancement of spectrum efficiency without the need for additional power\cite{oam_mode_division,oam_TalbotJSAC,oam_quasiFractalUCA,OAM_NFC_mine}. Moreover, OAM waves can maintain this orthogonality in line-of-sight (LOS) channels, making it more suitable for LOS transmissions than multiple-input and multiple-output (MIMO) systems relying on multipath propagation\cite{oam_MIMO}. Thus, OAM for multiplexing is a promising technology for the sixth-generation (6G) mobile communications\cite{6G_OAMin,6G_OAMin2}. Studies and experiments also validate that OAM technology can be employed in practical wireless communications\cite{OAM_Uncompressed_Video_Transmission,oam_multiplexing2,OAM_ExperimentalCapacityGain}. However, divergence and crosstalk of OAM waves severely decrease the
spectrum efficiency of OAM systems\cite{oam_hollow,OAM_analog_interfree}. Moreover, OAM communications are sensitive to transmission paths. The multipath environment can destroy the orthogonality of OAM waves and thus leading to more complex precoding, channel-estimation, and channel-equalization steps\cite{OAM_multipath_ZHL}. A reliable channel model with balanced accuracy and complexity is essential for further system analysis and for solving aforementioned problems. Although there have been some studies on solving the divergence and crosstalk problems of OAM, there have been limited channel modelings and measurements of OAM transmissions\cite{OAM_crosstalkMeasurements,MIMO_OAM_two-ray_propagation,UCA_OAM_Six-Ray_Canyon_Channel} to provide concise and reliable references for OAM-system design and standard construction.

\begin{figure*}[hb]
\centering
%\vspace{-10pt}
\includegraphics[scale=0.195]{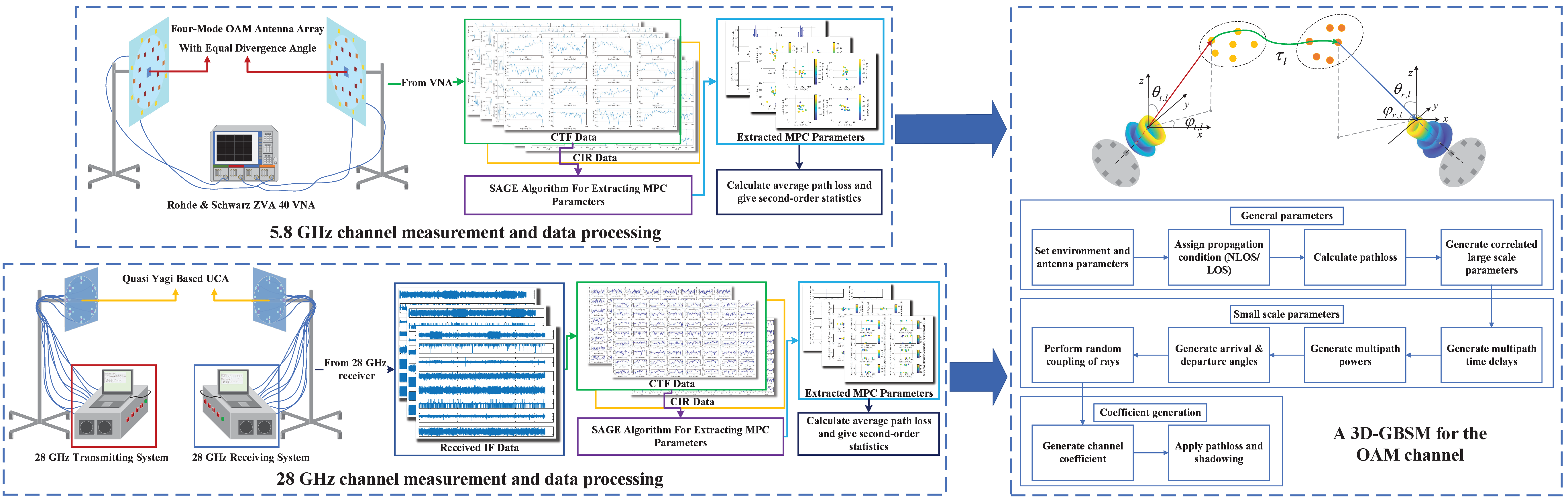}
\vspace{-5pt}
\caption{OAM channel measurement and channel modeling system model.}\label{fig:systemModel}
%\vspace{-5pt}
\end{figure*}

There are three main types of channel models, including parametric statistical channel models, deterministic channel models, and quasi-deterministic channel models. Statistical channel models\cite{Statistical_Characterization_kappaMu,Rician_MIMO,SimulationModelsRayleighRician,3DmmWaveStatisticalChannelModel} provide simple mathematical frameworks and statistical parameters to describe wireless channel characteristics, offering flexibility in modeling various conditions. However, as transmission frequencies increase and accuracy requirements grow, these models may oversimplify complex channel behaviors due to inherent assumptions. In 6G scenarios\cite{6G_OAMin2,3D_UAV_Rician}, statistical channel models become increasingly inaccurate in providing reliable space-time or angle-delay references for system analysis. They also struggle to describe the divergence, misalignment, and orthogonality of OAM signals. Deterministic channel models use methods like ray-tracing\cite{RayTracing_tunnels} and finite-difference time-domain (FDTD)\cite{UWB_onBody_FDTD_modeling,massiveMIMO_rayTrac_FDTD_modeling}, which describe channels using deterministic functions or algorithms. These models consider the physical aspects of the communication environment, such as propagation paths, obstacles, and antenna characteristics, enabling precise simulation and analysis in well-defined environments. As techniques like OAM-based wireless communications and reconfigurable intelligent surfaces (RIS)\cite{Intelligent_Surfaces,OAM-SWIPT,RIS,RIS_smartEnviroment} gain traction, deterministic channel models are increasingly useful for designing and optimizing antenna systems, beamforming techniques, and propagation mitigation strategies. However, they have limited applicability in dynamic or unpredictable scenarios and require accurate environmental parameters, which may be difficult to obtain in practice. Quasi-deterministic models, such as the geometric-based stochastic model (GBSM) and three-dimensional (3D) GBSM\cite{GBSM,GBSM_COST2100,GBSM_twinCluster,MassiveMIMOChannelMeasurements}, combine the advantages of statistical and deterministic models. They incorporate geometric information about the propagation environment, like the locations of scatterers and obstacles, to simulate deterministic propagation paths while introducing statistical variations to account for factors like fading and shadowing. This approach captures the stochastic nature of real-world channels, offering a general modeling framework and realistic simulation while maintaining computational efficiency. Therefore, in this paper, we propose a quasi-deterministic channel model for OAM channels.

The generation and reception of high-purity OAM beams are fundamental for OAM channel measurement and modeling. In lower frequency bands, a vector network analyzer (VNA) and high-purity OAM generation antennas can be used to measure OAM channels. However, at higher frequencies, such as in millimeter bands, RF equal-phase cables exhibit significant signal attenuation, causing received signals to fall below the VNA's sensitivity threshold. Thus, for OAM channel measurement in high-frequency bands, a high-power OAM beam generator with a robust amplifier and a highly sensitive OAM wireless signal receiver are essential. Using a uniform circular array (UCA) for OAM beam generation introduces stringent requirements for synchronizing RF channels across all UCA elements. This synchronization is crucial for maintaining phase coherence of the transmitted signals, ensuring the generation of highly pure OAM beams. Failure to maintain phase coherence compromises the purity of the OAM beams, leading to inaccuracies in OAM channel measurement and modeling. This is also one of the most significant problems to be solved. To address this major challenge, we developed an $8$-channel OAM generation device with synchronized RF channels operating at a central frequency of $28$ GHz, ensuring the integrity of the generated OAM beams. Paired with our $8$-channel millimeter-wave receiver, this setup facilitates accurate measurement of OAM channels.

%In this paper, we measure and model OAM channels in $5.8$ GHz and $28$ GHz bands. We first introduce the systems and the measurement scenarios. Then, we propose a 3D-GBSM for the OAM channel, which is modified from the twin-cluster-based 3D-GBSM\cite{GBSM_twinCluster,GBSM_COST2100} and can describe various stochastic properties of OAM channels, including the path loss, time-of-arrival (ToA), direction-of-departure (DoD), and direction-of-arrival (DoA) for each multipath component (MPC). Based on our proposed OAM channel model, the measurement data processing procedure is introduced. Next, we give the MPC parameters of the OAM 3D-GBSM according to the processed data in the given measurement environment. The second-order stochastic properties and the average path losses are also calculated. At last, we give the simulation channel generation scheme in the given measurement environment.

The main contributions of this paper are summarized as follows:
%\vspace{-5pt}
\begin{itemize}
\setlength{\itemsep}{0pt}
\setlength{\parsep}{0pt}
\setlength{\parskip}{0pt}
  \item [1)]
  \textbf{Development of 8-Channel OAM transmitting and receiving systems for $28$ GHz band:} This work develops an $8$-channel OAM generation device operating at a central frequency of 28 GHz, designed to maintain synchronized RF channels, thereby ensuring the integrity and purity of the generated OAM beams. A robust amplifier and a highly sensitive $8$-channel millimeter-wave receiver are developed to facilitate accurate OAM channel measurements at high frequencies.
  \item [2)]
  \textbf{Comprehensive Measurement and Modeling in 5.8 GHz and 28 GHz Bands:} This study measures and models OAM channels in both $5.8$ GHz and $28$ GHz bands, introducing the measurement systems and scenarios, processing measurement data, and calculating second-order stochastic properties and average path losses.
  \item [3)]
  \textbf{A new 3D-GBSM for the OAM channel:} This paper presents a modified 3D-GBSM for OAM channels, which describes various stochastic properties such as path loss, time-of-arrival (ToA), direction-of-departure (DoD), and direction-of-arrival (DoA) for each multipath component (MPC). This paper also provides a simulation channel generation scheme based on the measured data and proposed OAM channel model, aiding further analysis and optimization of OAM-based wireless communication systems.
\end{itemize}
%\vspace{-5pt}

%The rest of the paper is organized as follows. In the Results section, we first introduce the OAM channel sounder system, which includes antennas and a VNA for $5.8$ GHz channel measurement and the antenna, and the $8$-channel OAM transmitting and receiving systems for the $28$ GHz band. Next, we describe the channel measurement scenarios for both the $5.8$ GHz and $28$ GHz bands. Based on the measurement data, we provide the channel measurement data processing methods to extract MPC parameters and calculate the path loss. Following this, we present a 3D-GBSM for OAM channels. We then propose the steps for simulation channel generation. Finally, we analyze the channel measurement and simulation results at the end of the Results section. In the Methods section, we introduce the techniques for generating and detecting OAM beams and provide a brief overview of the MPC extraction method.

The rest of the paper is organized as follows. Section~\ref{sec:systemModel} gives the system model for OAM channel measurement and modeling. Section~\ref{sec:sounder} introduces the OAM channel sounder systems. Section~\ref{sec:scenario} describes the channel measurement scenarios for both $5.8$ GHz and $28$ GHz bands. The channel measurement data processing steps are introduced in Section~\ref{sec:dataProc}. Based on the processed data, a 3D-GBSM for OAM channels is developed in Section~\ref{sec:channelModel}. Section~\ref{sec:channelGen} proposes the steps for simulation channel generation. In Section~\ref{sec:results}, the channel measurement and simulation results are given. The conclusion is given in Section~\ref{sec:conclustion}.

$Notation:$ Matrices and vectors are denoted by the capital letters and the lowercase letters in bold, respectively. ``$\cdot$" represents the Hadamard product and ``$\otimes$" denotes the Kronecker product. The notations ``$(\cdot)^*$" and ``$(\cdot)^H$" denote the conjugation and Hermitian of a matrix or vector, respectively.

\section{System Model}\label{sec:systemModel}
%\begin{figure*}[htb]
%\centering
%%\vspace{-10pt}
%\includegraphics[scale=0.195]{pics//systemModel.eps}
%%\vspace{-15pt}
%\caption{OAM channel measurement and channel modeling system model.}\label{fig:systemModel}
%%\vspace{-5pt}
%\end{figure*}
This work can be mainly divided into three parts as shown in Fig.~\ref{fig:systemModel}, including $5.8$ GHz OAM channel measurement and data processing, $28$ GHz OAM channel measurement and data processing, and 3D-GBSM for OAM channel modeling. For channel measurement and data processing, we will introduce channel sounder systems, measurement scenarios, and measurement data processing. We use a common VNA and a pair of OAM antennas to measure the OAM channel at $5.8$ GHz. To address the significant signal attenuation, we develop a high-power OAM beam generator and a sensitive OAM wireless receiver for $28$ GHz channel data measurement. The measurement scenarios include indoor corridor LOS, through-wall non-line-of-sight (NLOS), and outdoor LOS scenarios at $5.8$ GHz. We also measure the OAM channel at 28 GHz in indoor corridor LOS and through-wall NLOS scenarios. Data processing involves using the space-alternating generalized expectation-maximization (SAGE) algorithm to extract MPC parameters, followed by second-order statistics and path loss calculations. Based on the measured data and calculated statistics, we model a 3D-GBSM for the OAM channel. In the following sections, we will introduce our work in detail.

\section{OAM Channel Sounder System}\label{sec:sounder}
\begin{figure*}[htbp]
\centering
%\vspace{-15pt}
\subfigure[OAM channel sounder at $5.8$ GHz.]{
\begin{minipage}{1\linewidth}
\centering
\includegraphics[scale=0.33]{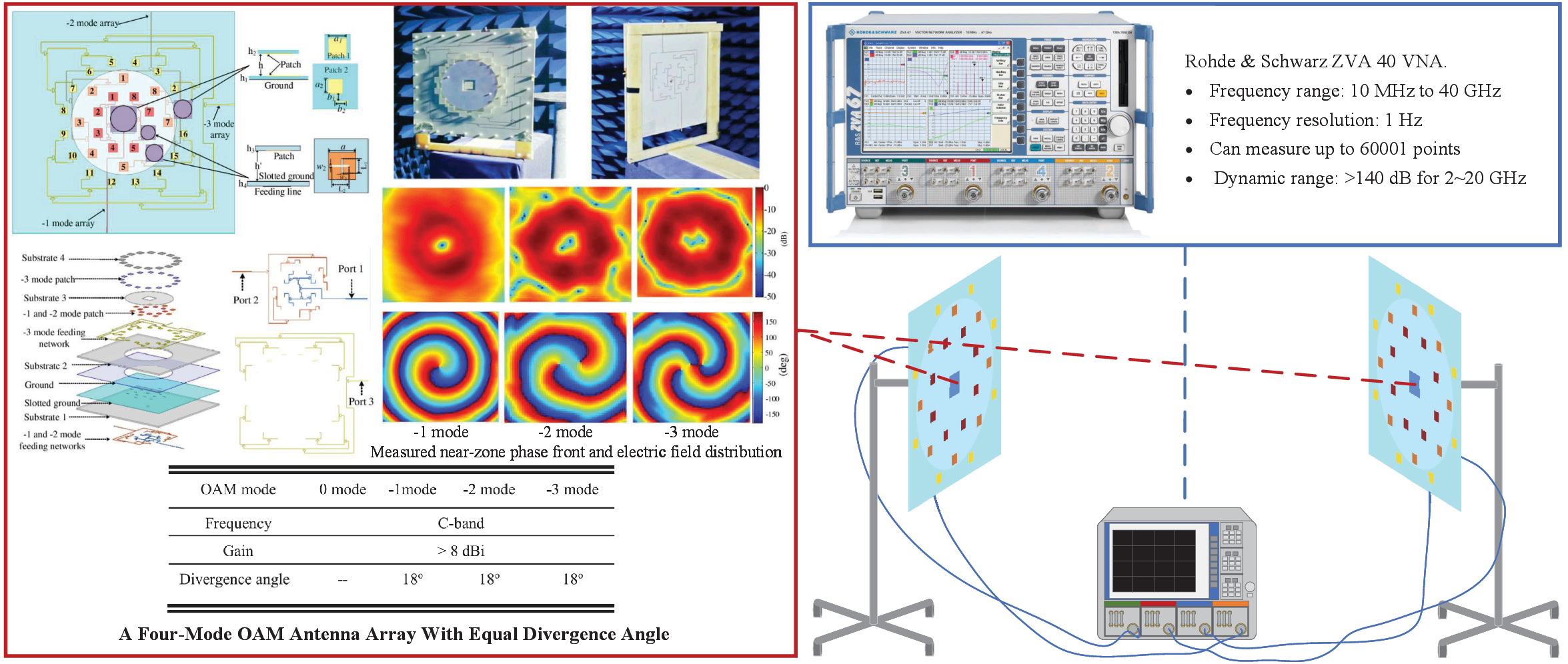}\label{fig:measurementFullPlot5.8G}
%\vspace{-5pt}
\end{minipage}
}\\
\subfigure[OAM channel sounder at $28$ GHz.]{
\begin{minipage}{1\linewidth}
\centering
\includegraphics[scale=0.33]{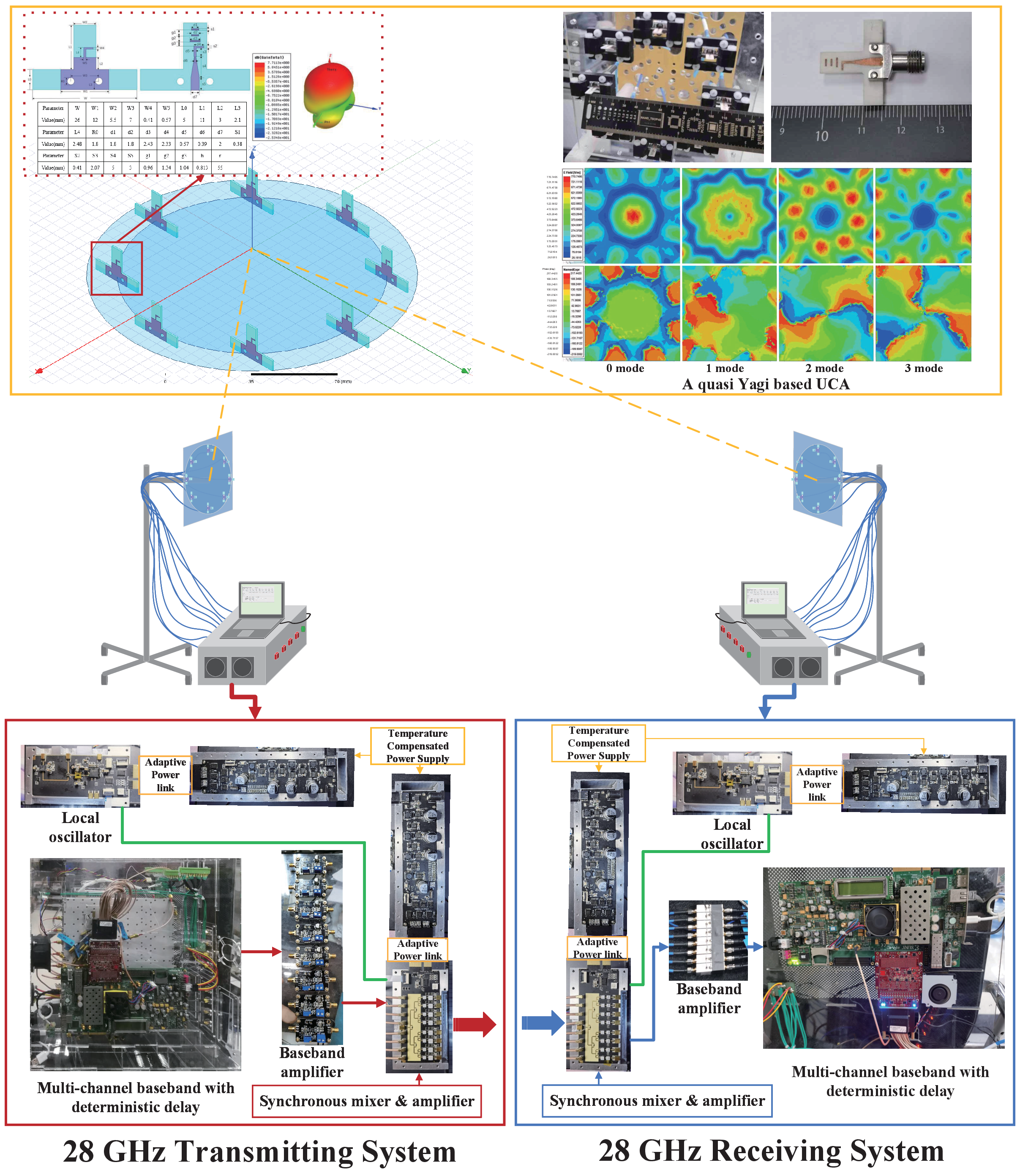}\label{fig:measurementFullPlot28G}
%\vspace{-5pt}
\end{minipage}
}
\centering
%\vspace{-10pt}
\caption{OAM channel sounder system.}\label{fig:measurementFullPlot}
%\vspace{-10pt}
\end{figure*}
As shown in Fig.~\ref{fig:measurementFullPlot} in page $4$, we use two different sounder systems to measure the OAM channels at $5.8$ GHz and $28$ GHz respectively. At $28$ GHz, we use two OAM antennas and a Rohde \& Schwarz ZVA $40$ VNA to measure the OAM channel. At $28$ GHz, since RF equal-phase cables exhibit significant signal attenuation, resulting in received signals falling below the sensitivity threshold of the VNA, we use a pair of UCAs along with our designed $28$ GHz $8$-chain transmitting and receiving systems to measure the OAM channel. The details are provided below.

\subsection{OAM Channel Sounder at 5.8 GHz}
Figure~\ref{fig:measurementFullPlot5.8G} illustrates the OAM channel sounder at $5.8$ GHz, which mainly consists of two OAM antennas and a VNA. The OAM generation and receiving antennas used in our measurement can transmit four-mode OAM waves with equal divergence angle\cite{oam_generation_EqualDivergenceAngle}. These antennas consist of a center patch antenna and three concentric UCAs to carry $0$, $-1$, $-2$, $-3$ OAM modes from center to outer rings, respectively. Each UCA is carefully optimized to control the different mode OAM waves to propagate along the same divergence angle as $18$ degrees. The antenna gains are above $8$ dBi for all modes. Measured results show that the four-mode OAM waves with equal divergence angle can be well generated. Rohde \& Schwarz ZVA $40$ VNA is used in $5.8$ GHz OAM channel measurement, which has a frequency range from $10$ MHz to $40$ GHz. Its frequency resolution is $1$ Hz and can measure up to $60001$ points. The dynamic range of this VNA is above $140$ dB for $2$ to $20$ GHz.

\subsection{OAM Channel Sounder at 28 GHz}
Figure~\ref{fig:measurementFullPlot28G} shows the OAM channel sounder operating at $28$ GHz. It primarily consists of two OAM antennas, custom $28$ GHz $8$-chain transmitting and receiving systems, and two laptops. We use a pair of UCAs, each with eight quasi-Yagi antennas, for the transmit and receive antennas. Each quasi-Yagi antenna has a gain of $7.7$ dBi and a bandwidth of around $10$ GHz. The antennas are arranged in a UCA with a radius of $55$ mm, using a Rogers $4003$ dielectric substrate with a constant of $3.55$ and a thickness of $0.8$ mm. These antennas support up to eight OAM modes simultaneously, from $-3$ to $4$.

At $28$ GHz, RF equal-phase cables cause significant signal attenuation, lowering the received signals below the VNA sensitivity threshold. To measure the OAM channel in high-frequency bands, we developed a high-power OAM beam generator and a sensitive OAM wireless receiver. The OAM beam generator includes a $28$ GHz $8$-chain transmitter, a $27$-$32$ GHz local oscillator, a hot-swappable power supply, and an integrated debugger sub-board. The calibrated signal strengths of the transmitter's eight chains range from $-0.24$ to $-0.02$ dBm, with a maximum difference of $0.22$ dB. Similarly, the OAM receiver consists of an $8$-chain receiver, local oscillator, power supply, and debugger sub-board. Air interface testing verified the transmitter and receiver's functionality, showing signal strengths from $-46.18$ to $-47.46$ dBm across the eight chains, with a maximum difference of $1.28$ dB, well within the $3$ dB limit. Two laptops with our developed drivers and GUIs are used to set measurement parameters and record data.

Key system parameters include:
%\vspace{-5pt}
\begin{itemize}
\setlength{\itemsep}{0pt}
\setlength{\parsep}{0pt}
\setlength{\parskip}{0pt}
  \item [1)]
  ADC and DAC accuracies: $16$ bits
  \item [2)]
  ADC sampling frequency: $250$ MHz
  \item [3)]
  DAC sampling frequency: $510$ MHz
  \item [4)]
  Baseband gains of transmit and receive power amplifiers: $32$ and $31$ dB, respectively
  \item [5)]
  RF gains of transmit and receive power amplifiers: $21$ and $13$ dB, respectively
  \item [6)]
  Quantization accuracy: $16$ bits
  \item [7)]
  Clock accuracy: $0.1$ ppm
\end{itemize}
%\vspace{-5pt}

More specifically, the $28$ GHz $8$-chain transmitter, developed by our research group, generates OAM beams for channel measurements. It has $8$ synchronous RF channels operating at $28$ GHz, interfacing with $8$ UCA elements to generate multiple OAM modes. The transmitter integrates mixing and amplification functions and uses embedded sensors for real-time monitoring of parameters such as voltage, current, and temperature. Digital power management, PC-based monitoring, and active heat dissipation are also included. Achieving multi-channel synchronization at $28$ GHz is very challenging. Hence, we use a hybrid compensation approach combining digital and analog techniques to ensure channel coherence and OAM beam purity. Embedded processors enable real-time adjustments based on sensor data, and an integrated amplitude-phase compensation IP core adjusts phase and amplitude using calibration data. The $28$ GHz $8$-chain receiver is similar to the transmitter, with mixing and amplification, advanced synchronization, and real-time monitoring via sensors. It also includes digital power management, PC monitoring, and active heat dissipation. The $27$-$32$ GHz local oscillator features a low phase noise phase-locked loop and a two-stage amplifier. It includes an external reference signal input and a built-in $100$ MHz reference clock, outputting signals in the $27$-$32$ GHz band. Configurable via software, it supports simultaneous output of synchronous baseband signals from $62.6$ MHz to $8$ GHz, with dual output ports for different signal amplitudes and active heat dissipation. The integrated debugger sub-board has USB to RS$232$ and SWIM interfaces, enabling independent debugging and downloading of the single-chip microcomputer on the millimeter-wave sub-board without powering the entire board.

\section{Channel Measurement Scenarios}\label{sec:scenario}
\begin{figure}[htbp]
\centering
%\vspace{-15pt}
\subfigure[Indoor corridor LOS scenario.]{
\begin{minipage}{1\linewidth}
\centering
\includegraphics[scale=0.32]{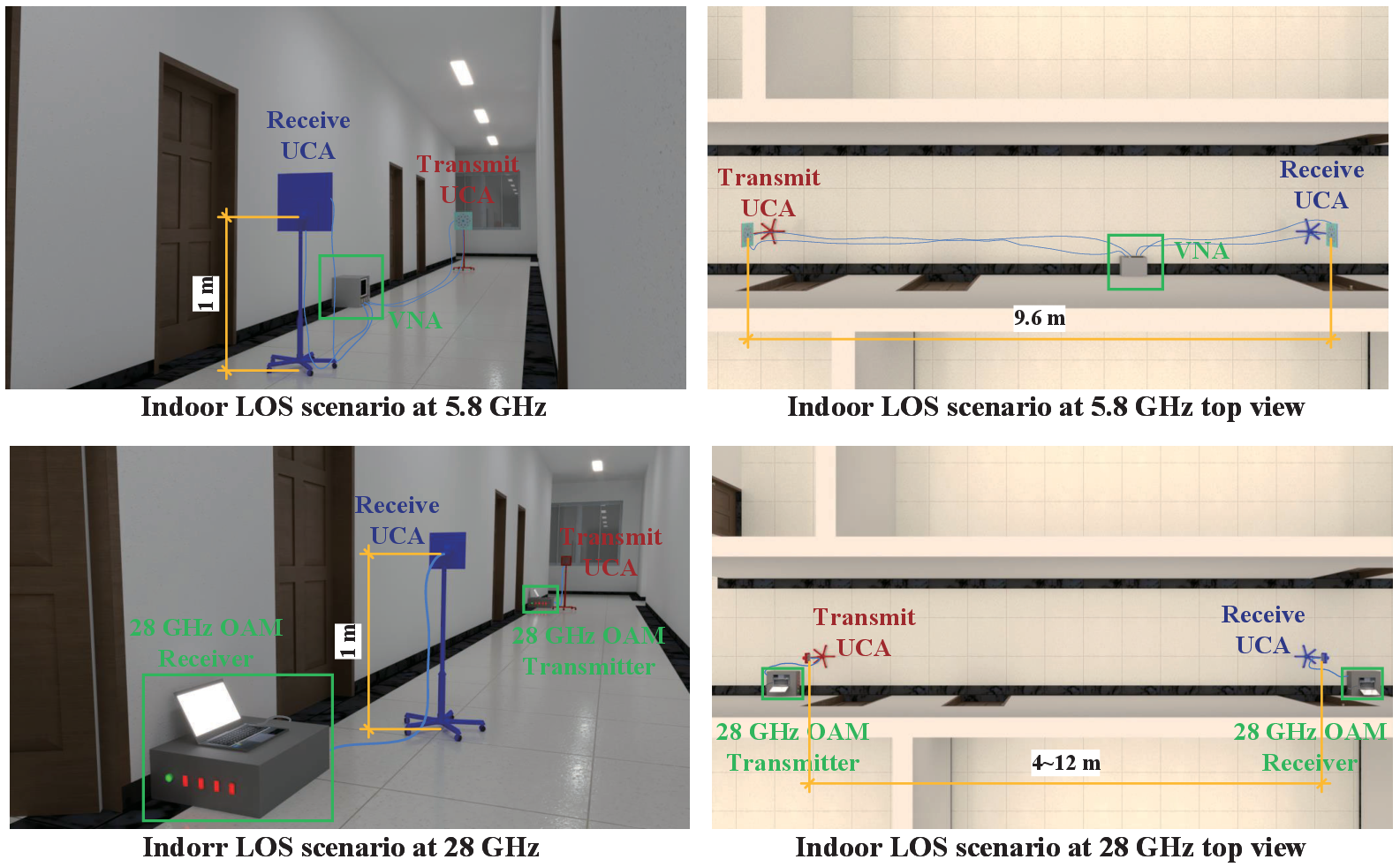}\label{fig:scenarioIndoorLOS}
%\vspace{-5pt}
\end{minipage}
}\\
\subfigure[Indoor through-wall scenario.]{
\begin{minipage}{1\linewidth}
\centering
\includegraphics[scale=0.32]{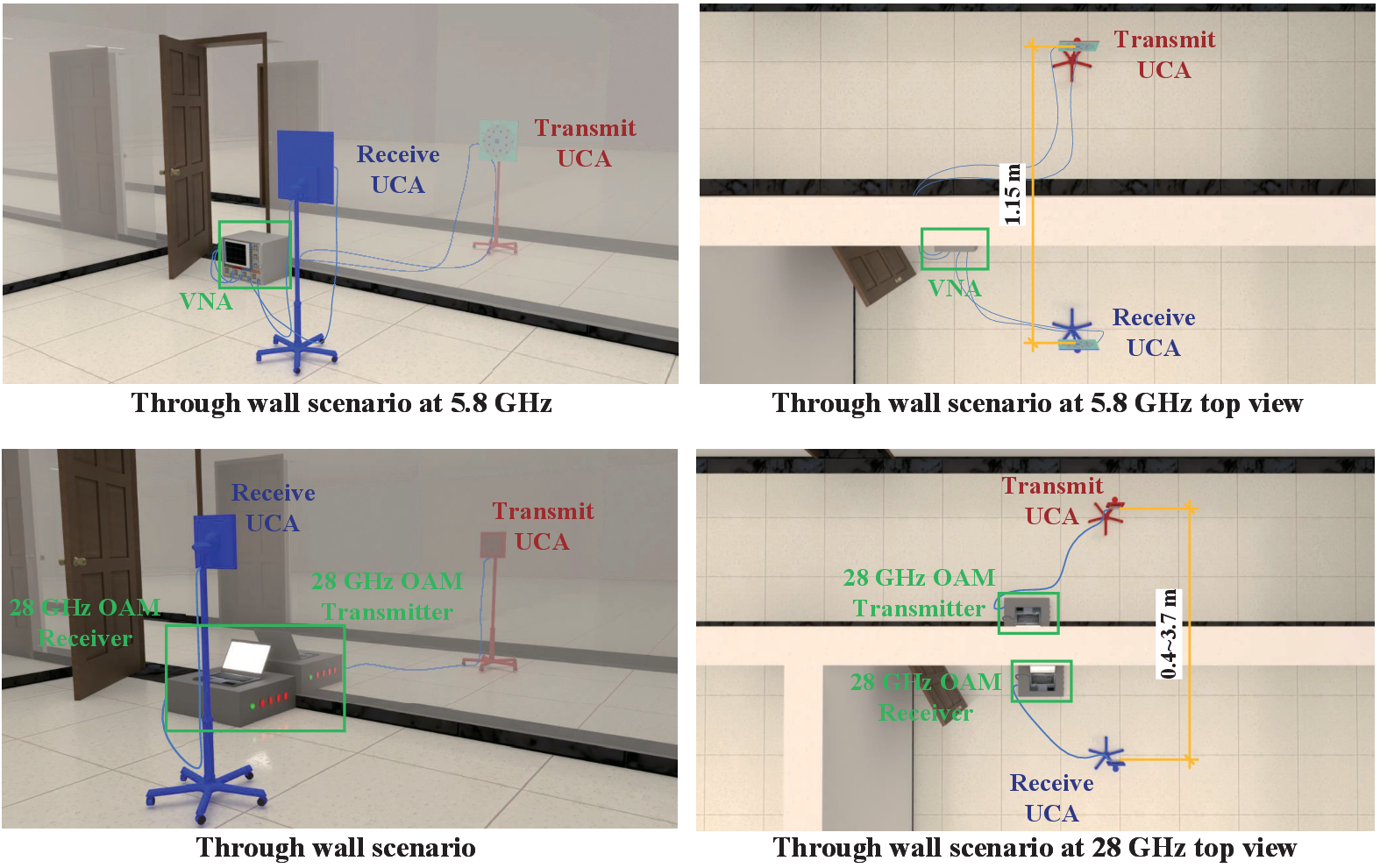}\label{fig:scenarioIndoorThroWall}
%\vspace{-5pt}
\end{minipage}
}\\
\subfigure[Outdoor LOS scenario.]{
\begin{minipage}{1\linewidth}
\centering
\includegraphics[scale=0.32]{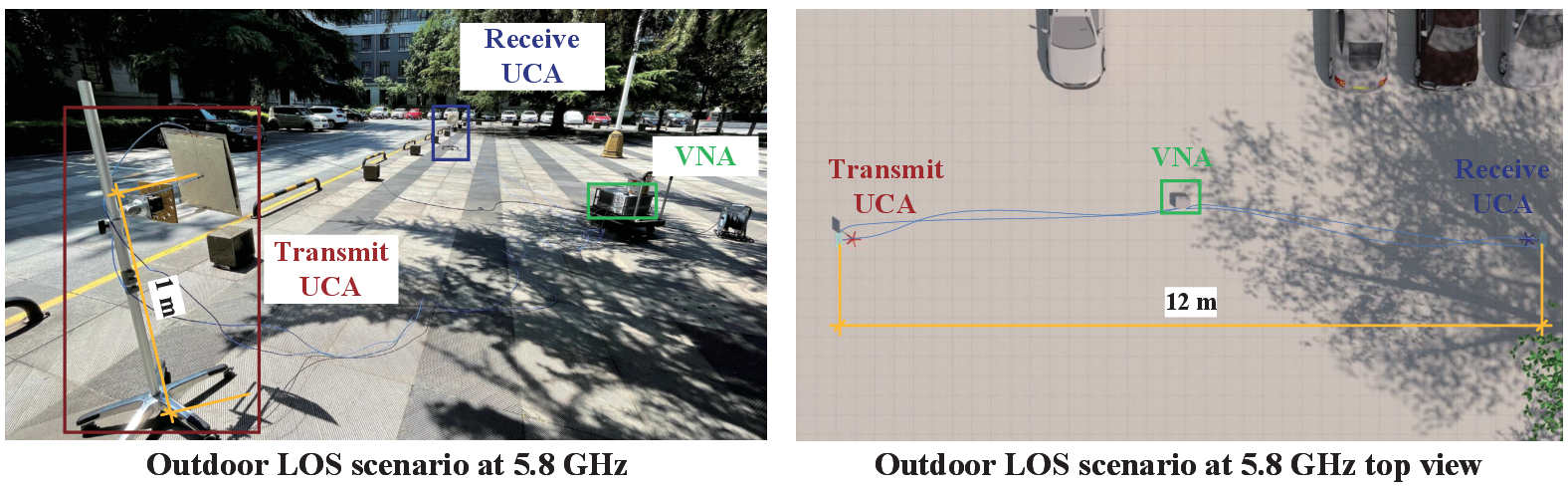}\label{fig:scenarioOutdoorLOS}
%\vspace{-5pt}
\end{minipage}
}
\centering
%\vspace{-10pt}
\caption{Measurement Scenarios.}\label{fig:scenario}
%\vspace{-10pt}
\end{figure}
The modeling scenarios include indoor corridor LOS, through-wall NLOS, and outdoor LOS scenarios at $5.8$ GHz. We also measure the OAM channel at $28$ GHz in indoor corridor LOS and through-wall NLOS scenarios. The corresponding channel data measurement schematic diagrams are shown in Fig.~\ref{fig:scenario}. Detailed scenario descriptions are presented as below.

\textbf{Indoor corridor LOS scenario:} As depicted in Fig.~\ref{fig:scenarioIndoorLOS}, the corridor comprises cement walls on both sides, a marble floor, and a foam board ceiling. With a width of $2.4$ meters and a height of $3.5$ meters, it provides a controlled environment for channel measurement. The distance between the transmitter and receiver is set as $9.6$ m at $5.8$ GHz and ranges from $4$ to $12$ meters at $28$ GHz, with the transmitter positioned at distances of $0.8$ meters and $1.6$ meters from each side wall. Antennas for transmitters and receivers at both $5.8$ and $28$ GHz have a height of $1$ meter. When operating at a central frequency of $5.8$ GHz, the bandwidth is set at $100$ MHz, utilizing $51$ sweeping points. At a higher frequency of $28$ GHz, a bandwidth of $1$ GHz is utilized. The measurement scheme includes $20$ sweeping points for the $4$ meter distance, $50$ sweeping points for distances of $6$, $8$, and $10$ meters, and $60$ sweeping points for the $12$ meter distance. %This scenario encompasses multipath propagation phenomena, which may significantly influence signal propagation within the environment.

\textbf{Indoor through-wall scenario:} In this scenario, a concrete wall with a thickness of $30$ cm is used as the propagation obstacle as in Fig.~\ref{fig:scenarioIndoorThroWall}. The distance between transmitter and receiver antennas is $1.15$ m at $5.8$ GHz and varies from $0.4$ to $3.7$ meters at $28$ GHz. Antennas are positioned at a height of $1$ meter. Operating at $5.8$ GHz, the bandwidth is set to $100$ MHz. A consistent measurement approach utilizing $51$ sweeping points is adopted. For the higher frequency of $28$ GHz, the bandwidth is set as $1$ GHz, with $30$ sweeping points applied across all distances.

\textbf{Outdoor LOS scenario:} As depicted in Fig.~\ref{fig:scenarioOutdoorLOS}, channel measurements in an outdoor environment at $5.8$ GHz are conducted over distance of $12$ m. The central frequency is at $5.8$ GHz, with a bandwidth of $100$ MHz. A total of $51$ sweeping points are measured in this scenario.

\begin{figure*}[htbp]
%\hrulefill
\setcounter{equation}{4}
\begin{align}
PL^{\rm Fit}_{n_r,n_t} = A - B\lg\left(\left|J_{m_{n_t}}\left(Cf^{\rm GHz}\frac{R_r}{\sqrt{R_r^2 + d^2}}\right)\right|\left|J_{m_{n_r}}\left(Cf^{\rm GHz}\frac{R_t}{\sqrt{R_t^2 + d^2}}\right)\right|\right)+D\lg(d) + E\lg(f^{\rm GHz}).\label{eq:PlFit}
\end{align}
\hrulefill
\end{figure*}
\section{Channel Measurement Data Processing}\label{sec:dataProc}
%\begin{figure}[htbp]
%\centering
%%\vspace{-15pt}
%\subfigure[Data analyses at $5.8$ GHz.]{
%\begin{minipage}{1\linewidth}
%\centering
%\includegraphics[scale=0.3]{pics//dataAnalyses5.8G.eps}\label{fig:dataAnalyses5.8G}
%%\vspace{-5pt}
%\end{minipage}
%}\\
%\subfigure[Data analyses at $28$ GHz.]{
%\begin{minipage}{1\linewidth}
%\centering
%\includegraphics[scale=0.3]{pics//dataAnalyses28G.eps}\label{fig:dataAnalyses28G}
%%\vspace{-5pt}
%\end{minipage}
%}\\
%\subfigure[SAGE algorithm flow chart.]{
%\begin{minipage}{1\linewidth}
%\centering
%\includegraphics[scale=0.57]{pics//SageFlowChart.eps}\label{fig:SageFlowChart}
%%\vspace{-5pt}
%\end{minipage}
%}
%\centering
%%\vspace{-10pt}
%\caption{OAM channel measurement data analyses.}
%%\vspace{-10pt}
%\end{figure}
The measurement data processing procedure is shown in Fig.~\ref{fig:systemModel}. Main steps are similar for data analyzing at both $5.8$ GHz and $28$ GHz, including the MPC extraction process, second-order statistic calculations, and calculating the average path loss. The difference is that for OAM channel at $5.8$ GHz, the input data are CTFs directly output from the VNA. However, for OAM channel at $28$ GHz, CTFs need to be calculated from the received IF data of the $28$ GHz received system. Detailed steps are given in the following of this part.

\subsection{SAGE Algorithm for Estimating Channel Parameters}
\begin{figure}[htbp]
\centering
%\vspace{-10pt}
\includegraphics[scale=0.5]{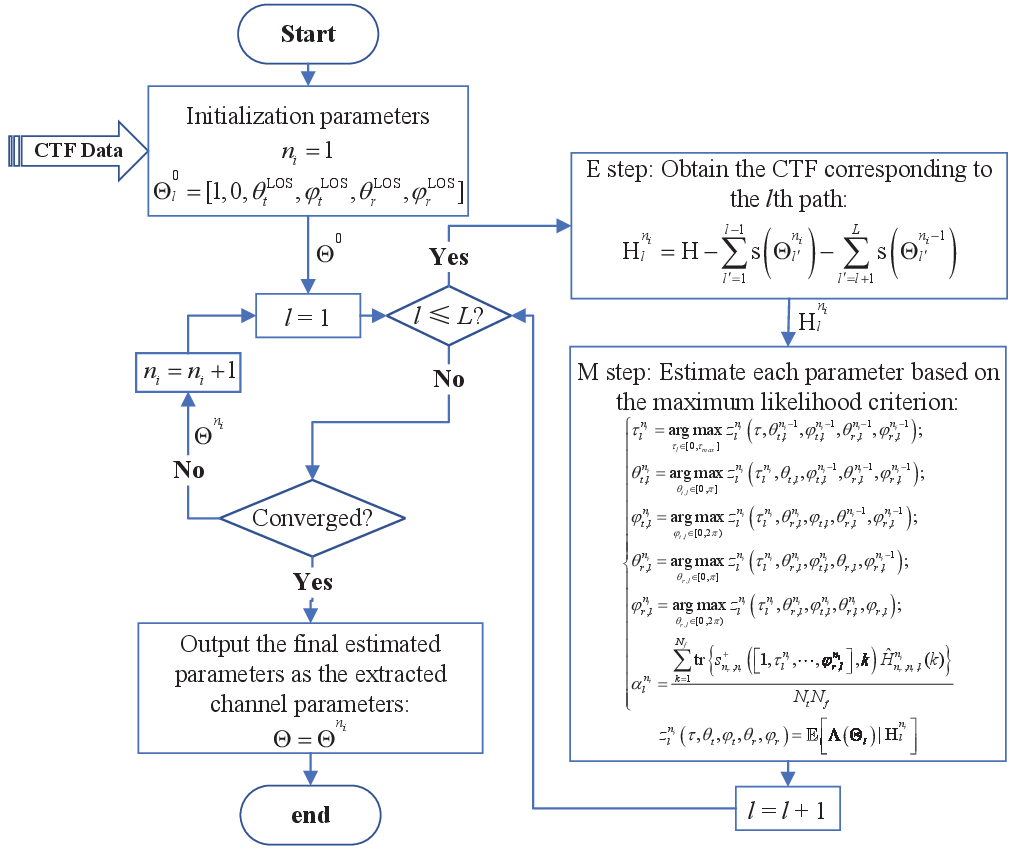}
%\vspace{-15pt}
\caption{SAGE algorithm flow chart.} \label{fig:SageFlowChart}
%\vspace{-5pt}
\end{figure}
Based on the CTFs recorded by the sounders, we use the SAGE algorithm to estimate the MPC parameters. These parameters include complex amplitude, delay, elevation angle-of-arrival (EoA), elevation angle-of-departure (EoD), azimuth angle-of-arrival (AoA), and azimuth angle-of-departure(AoD) of the $l$th MPC, denoted by $\alpha_l$, $\tau_l$, $\theta_{t,l}$, $\varphi_{t,l}$, $\theta_{r,l}$, and $\varphi_{r,l}$, respectively. The main steps of SAGE algorithm is given in Fig.~\ref{fig:SageFlowChart}, including initialization, E step, and M step. After convergence or after enough numbers of iterations, the estimated parameters will converge to stable values. The stable extracted MPC parameters are denoted by $\hat{\boldsymbol{\mathrm \Theta}}=\hat{\boldsymbol{\mathrm \Theta}}^{N_i}$, where $N_i$ is the maximum iteration times.

\subsection{MPC Parameters Processing}
Based on the MPC parameters estimated by SAGE algorithm, we can give the second-order statistics for every parameter. The root mean square (RMS) delay spread (DS), denoted by $\tau_{\rm RMS}$, is used to describe the dispersion of the power delay profile and is given as follows:
\setcounter{equation}{0}
\begin{align}
\tau_{\rm RMS}=\sqrt{\frac{\sum_l^L\left|\hat{\alpha}_l\right|^2\hat{\tau}_l^2}{\sum_l^L\left|\hat{\alpha}_l\right|^2}
-\left(\frac{\sum_l^L\left|\hat{\alpha}_l\right|^2\hat{\tau}_l}{\sum_l^L\left|\hat{\alpha}_l\right|}\right)^2}.\label{eq:RMSDS}
\end{align}
Similarly, the root mean square (RMS) angle spread (AS) for the transmit elevation angle, the transmit azimuth angle, the receive elevation angle, and the receive azimuth angle, denoted by $\theta_{t,RMS}$, $\varphi_{t,RMS}$, $\theta_{r,RMS}$, and $\theta_{r,RMS}$, respectively, can be given as follows:
\begin{align}
\begin{cases}
\theta_{t,\rm RMS}=\sqrt{\frac{\sum_l^L\left|\hat{\alpha}_l\right|^2\hat{\theta}_{t,l}^2}{\sum_l^L\left|\hat{\alpha}_l\right|^2}
-\left(\frac{\sum_l^L\left|\hat{\alpha}_l\right|^2\hat{\theta}_{t,l}}{\sum_l^L\left|\hat{\alpha}_l\right|}\right)^2};\\
\varphi_{t,\rm RMS}=\sqrt{\frac{\sum_l^L\left|\hat{\alpha}_l\right|^2\hat{\varphi}_{t,l}^2}{\sum_l^L\left|\hat{\alpha}_l\right|^2}
-\left(\frac{\sum_l^L\left|\hat{\alpha}_l\right|^2\hat{\varphi}_{t,l}}{\sum_l^L\left|\hat{\alpha}_l\right|}\right)^2};\\
\theta_{r,\rm RMS}=\sqrt{\frac{\sum_l^L\left|\hat{\alpha}_l\right|^2\hat{\theta}_{r,l}^2}{\sum_l^L\left|\hat{\alpha}_l\right|^2}
-\left(\frac{\sum_l^L\left|\hat{\alpha}_l\right|^2\hat{\theta}_{r,l}}{\sum_l^L\left|\hat{\alpha}_l\right|}\right)^2};\\
\varphi_{r,\rm RMS}=\sqrt{\frac{\sum_l^L\left|\hat{\alpha}_l\right|^2\hat{\varphi}_{r,l}^2}{\sum_l^L\left|\hat{\alpha}_l\right|^2}
-\left(\frac{\sum_l^L\left|\hat{\alpha}_l\right|^2\hat{\varphi}_{r,l}}{\sum_l^L\left|\hat{\alpha}_l\right|}\right)^2}.\label{eq:RMSAS}
\end{cases}
\end{align}

Besides, the correlation coefficient between the $n_1$th and the $n_2$th OAM-mode channels can be given as follows:
\begin{align}
\rho_{n_1,n_2} = \frac{1}{N_f}\sum_{k=1}^{N_f}\frac{\mathbb{E}\left\{H_{n_1,n_1}(k)H_{n_2,n_2}^*(k)\right\}}{\sqrt{\mathbb{E}\left\{\left|H_{n_1,n_1}(k)\right|^2\right\}\mathbb{E}\left\{\left|H_{n_2,n_2}(k)\right|^2\right\}}},
\end{align}
where ``$\mathbb{E}\{\cdot\}$" is the expectation operation.

\subsection{Calculate the Average Path Loss}
Utilizing the measured CTF, the received power can be determined for each OAM mode. Subsequently, by incorporating the transmit and receive gain, the path loss can be calculated. Following this, a path loss calculation formula can be fitted for each scenario based on the path losses observed for various OAM modes at different transceiver distances. The theoretical OAM received signal of the $n_t$th OAM mode at a point with its spherical coordinate denoted by $(\rho,\theta,\varphi)$ can be given as follows:
\begin{align}
E_{n_t}(\rho,\theta,\varphi)\hspace{-0.1cm}=\hspace{-0.1cm}\sqrt{N}\frac{j^{n_t}c}{4\pi f}J_{m_{n_t}}\hspace{-0.1cm}\left(\hspace{-0.1cm}\frac{2\pi f}{c}R_t\sin\theta\hspace{-0.1cm}\right)\hspace{-0.1cm}\frac{e^{-j\frac{2\pi f\rho}{c}}}{\rho}e^{jm_{n_t}\varphi},
\end{align}
where $J_\alpha(x)$ denotes Bessel function of the first kind of order $\alpha$. Therefore, the fitting formula for path loss in dB from the $n_t$th transmit OAM mode to the $n_r$th receive OAM mode, denoted by $PL^{\rm Fit}_{n_r,n_t}$, can be given as in Eq.~\eqref{eq:PlFit}, %follows:
%\begin{align}
%PL^{\rm Fit}_{n_r,n_t} = A - B\lg\left(\left|J_{m_{n_t}}\left(Cf^{\rm GHz}\frac{R_r}{\sqrt{R_r^2 + d^2}}\right)\right|\left|J_{m_{n_r}}\left(Cf^{\rm GHz}\frac{R_t}{\sqrt{R_t^2 + d^2}}\right)\right|\right)+D\lg(d) + E\lg(f^{\rm GHz}),\label{eq:PlFit}
%\end{align}
where $A$, $B$, $C$, $D$ and $E$ are fitting parameters, $f^{\rm GHz}$ denotes $f$ in GHz, and $d$ denotes the distance between transmitter and receiver centers. We employ two Bessel functions in Eq.~\eqref{eq:PlFit} to characterize the beam gains of the two OAM modes at the transmitter and receiver, respectively. For the transmit OAM mode, the beam gain at the receiving position can be determined by considering the radius of the receive antenna and the transmitting distance when the transmit and receive antennas are aligned with each other. Similarly, the beam gain of the receive OAM mode at the transmitting position can be determined based on the radius of the transmit antenna and the transmitting distance. This approach allows for a comprehensive evaluation of beam gains for both transmitting and receiving OAM modes, considering the specific configurations and distances involved in the communication setup. %For a more simplified case where the transmitted and received radii are the same, the fitting formula for path loss in dB form the $n_t$th transmit OAM mode to the same received OAM mode can be given as follows:
%%\begin{align}
%%PL^{\rm Fit}_{n_r,n_t} = A + 2B\lg\left(\left|J_{m_{n_t}}\left(\frac{2\pi f}{c}\frac{R_r}{\sqrt{R_r^2 + d^2}}\right)\right|\right) + C\lg(d) + D\lg(f^{\rm GHz}).\label{eq:PlFit2}
%%\end{align}
%\begin{align}
%PL^{\rm Fit}_{n_t,n_t} = A - 2B\lg\left(\left|J_{m_{n_t}}\left(Cf^{\rm GHz}\frac{R_r}{\sqrt{R_r^2 + d^2}}\right)\right|\right)+D\lg(d) + E\lg(f^{\rm GHz}).\label{eq:PlFit2}
%\end{align}

\textbf{Log-normal shadowing model calculation:}

We use log-normal shadowing model to describe shadow fading. In this model, the shadow fading (dB form), denoted by $\psi$, is a normally distributed random variable. Its probability density is given as follows:
\setcounter{equation}{5}
\begin{align}
p(\psi)=\frac{10/\ln10}{\sqrt{2\pi}\sigma_{\psi}\psi}\exp{\left[-\frac{(10\lg\psi)^2}{2\sigma_{\psi}^2}\right]},
\end{align}
where $\sigma_{\psi}^2$ denotes the variance of the shadow fading in dB form.

\section{OAM Channel Model}\label{sec:channelModel}

\begin{table}[ht]
\centering
\renewcommand{\arraystretch}{1.3} % Default value: 1
\begin{tabular}{|c|c|} % <-- Alignments: 1st column left, 2nd middle and 3rd right, with vertical lines in between
\hline
\textbf{Parameter name} & \textbf{Definition}\\
\hline
$\alpha_l$, $\tau_l$ & the path gain and ToA of the $l$th MPC\\
\hline
$\boldsymbol{\mathrm \Omega}_{r,l}$, $\boldsymbol{\mathrm \Omega}_{t,l}$ & the DoD and DoA of the $l$th MPC\\
%\hline
%$c_{r,n_r}\left(\boldsymbol{\mathrm \Omega}_{r,l}\right)$, $c_{t,n_t}\left(\boldsymbol{\mathrm \Omega}_{r,l}\right)$ & the array antenna patterns of the $n_r$th receive array and the $n_t$th transmit array\\
\hline
$c_{r,n_r}\left(\boldsymbol{\mathrm \Omega}_{r,l}\right)$, & the array antenna patterns of the $n_r$th\\
$c_{t,n_t}\left(\boldsymbol{\mathrm \Omega}_{r,l}\right)$ & receive array and the $n_t$th transmit array\\
\hline
$a_{r,n_r}\left(\theta_{r,l},\varphi_{r,l}\right)$, & the phases of the $l$th MPC of the $n_r$th receive\\
$a_{t,n_t}\left(\theta_{t,l},\varphi_{t,l}\right)$ & OAM-mode and the $n_t$th transmit OAM-mode\\
\hline
$\theta_{r,l}$, $\theta_{t,l}$ & the EoA and EoD for the $l$th MPC\\
\hline
$\varphi_{r,l}$, $\varphi_{t,l}$ & the AoA and AoD for the $l$th MPC\\
%\hline
%      $\boldsymbol{\mathrm p}_{r,n_r}$, $\boldsymbol{\mathrm p}_{t,n_t}$ & the coordinates of the $n_r$th receive antenna and the $n_t$th transmit antenna\\
\hline
\end{tabular}
\vspace{5pt}
\caption{Definitions of GBSM Parameters.}\label{tab:GBSM_para}
\end{table}

In this part, we propose the channel model for OAM channels. For the $n_t$th transmit OAM mdoe and the $n_r$th receive OAM mode, the channel transfer function (CTF) of frequency $f$, denoted by $H_{n_r,n_t}(f)$, is given as follows:
%\begin{align}
%H_{n_r,n_t}(f)=\sum_{l=1}^L&\alpha_lc_{r,n_r}\left(\boldsymbol{\mathrm \Omega}_{r,l}\right)c^*_{t,n_t}\left(\boldsymbol{\mathrm \Omega}_{t,l}\right)\exp{\left[ja_{t,n_t}\left(\theta_{t,l},\varphi_{t,l}\right)-ja_{r,n_r}\left(\theta_{r,l},\varphi_{r,l}\right)\right]}\exp{\left(-j2\pi f\tau_l\right)},
%\label{eq:CTF}
%\end{align}
\begin{align}
H_{n_r,n_t}(f)=&\sum_{l=1}^L\alpha_lc_{r,n_r}\left(\boldsymbol{\mathrm \Omega}_{r,l}\right)c^*_{t,n_t}\left(\boldsymbol{\mathrm \Omega}_{t,l}\right)\nonumber\\
&\quad\exp{\left[ja_{t,n_t}\left(\theta_{t,l},\varphi_{t,l}\right)-ja_{r,n_r}\left(\theta_{r,l},\varphi_{r,l}\right)\right]}\nonumber\\
&\quad\quad\exp{\left(-j2\pi f\tau_l\right)},
\label{eq:CTF}
\end{align}
where $L$ denotes the number of MPCs and other parameters, including the path gain, ToA, DoD, DoA, EoA, EoD, AoA, and AoD of the $l$th MPC, are listed in Table~\ref{tab:GBSM_para} for more clear clarification. In Eq.~\eqref{eq:CTF}, $\boldsymbol{\mathrm \Omega}_{t,l}$ and $\boldsymbol{\mathrm \Omega}_{r,l}$ can be determined by $\theta_{t,l}$ and $\varphi_{t,l}$, $\theta_{r,l}$ and $\varphi_{r,l}$ respectively, as follows:
\begin{align}
\begin{cases}
\boldsymbol{\mathrm \Omega}_{t,l} = \left[\cos\varphi_{t,l}\sin\theta_{t,l},
\sin\varphi_{t,l}\sin\theta_{t,l},\cos\theta_{t,l}\right]^T;\\
\boldsymbol{\mathrm \Omega}_{r,l} = \left[\cos\varphi_{r,l}\sin\theta_{r,l},
\sin\varphi_{r,l}\sin\theta_{r,l},\cos\theta_{r,l}\right]^T.
\end{cases}
\end{align}
Furthermore, $c_{t,n_t}\left(\boldsymbol{\mathrm \Omega}_{t,l}\right)$ and $c_{r,n_r}\left(\boldsymbol{\mathrm \Omega}_{r,l}\right)$ are expressed using $\boldsymbol{\mathrm \Omega}_{t,l}$ and $\boldsymbol{\mathrm \Omega}_{r,l}$ as follows:
\begin{align}
\begin{cases}
c_{t,n_t}\left(\boldsymbol{\mathrm \Omega}_{t,l}\right) = \exp{\left(j2\pi\frac{f}{c}\boldsymbol{\mathrm p}_{t,n_t}\boldsymbol{\mathrm \Omega}_{t,l}\right)};\\
c_{r,n_r}\left(\boldsymbol{\mathrm \Omega}_{r,l}\right) = \exp{\left(j2\pi\frac{f}{c}\boldsymbol{\mathrm p}_{r,n_r}\boldsymbol{\mathrm \Omega}_{r,l}\right)},
\end{cases}
\end{align}
where $c$ represents the light speed, $\boldsymbol{\mathrm p}_{r,n_r}\in \mathbb{R}^{1\times 3}$ denotes the coordinate of the $n_r$th receive antenna, and $\boldsymbol{\mathrm p}_{t,n_t}\in \mathbb{R}^{1\times 3}$ denotes the coordinate of the $n_t$th transmit antenna.
%Also, $a_{t,n_t}$ and $a_{r,n_r}$ can be determined by the orientations and radii of the OAM antennas as well as the carrier frequency as follows:
%\begin{align}
%\begin{cases}
%a_{t,n_t}\left(\theta,\varphi\right) = {\rm rot}\left(\hat{a}_{f,R_t,n_t}\left(\theta,\varphi\right),[\phi_{t,x},\phi_{t,y},\phi_{t,z}]\right);\nonumber\\
%a_{r,n_r}\left(\theta,\varphi\right) = {\rm rot}\left(\hat{a}_{f,R_r,n_r}\left(\theta,\varphi\right),[\phi_{t,x},\phi_{t,y},\phi_{t,z}]\right),
%\end{cases}
%\end{align}
%where $R_t$ and $R_r$ denote the transmit and receive OAM antenna radii, respectively; $\hat{a}_{f,R_t,n_t}$ and $\hat{a}_{f,R_t,n_r}$ denote the OAM antenna pattern of the $n_t$ and $n_r$ modes, respectively; $[\phi_{t,x},\phi_{t,y},\phi_{t,z}]$ and $[\phi_{r,x},\phi_{r,y},\phi_{r,z}]$ denote the rotation angles around the $x$-, $y$-, and $z$- axes according to the right-hand spiral law for the transmit and receive OAM antennas, respectively. $\hat{a}_{f,R_t,m}\left(\theta,\varphi\right)$ can be expressed as follows:
%\begin{align}
%&\hat{a}_{f,R_t,m}\left(\theta,\varphi\right) \\ &\ \ =
%\begin{cases}
%J_m\left(\frac{4\pi cR_t}{f}\sin\theta\right)\exp{\left(jm\varphi\right)},\ {\rm for}\ \theta\in[0,pi/2];\\
%\hspace{2cm}0\hspace{2cm},\ {\rm for\ otherwise},
%\end{cases}
%\end{align}
%where $m=n_t\ {\rm or}\ n_r$.
Also in Eq.~\eqref{eq:CTF}, $a_{t,n_t}$ and $a_{r,n_r}$ can be determined by the orientations of the OAM antennas as follows:
\begin{align}
\begin{cases}
a_{t,n_t}\left(\theta,\varphi\right) = {\rm rot}\left(\hat{a}_{n_t}\left(\theta,\varphi\right),[\phi^{\rm Tx}_{x},\phi^{\rm Tx}_{y},\phi^{\rm Tx}_{z}]\right);\\
a_{r,n_r}\left(\theta,\varphi\right) = {\rm rot}\left(\hat{a}_{n_r}\left(\theta,\varphi\right),[\phi^{\rm Rx}_{x},\phi^{\rm Rx}_{y},\phi^{\rm Rx}_{z}]\right),
\label{eq:OAM_phase}
\end{cases}
\end{align}
where $\hat{a}_{n_t}$ and $\hat{a}_{n_r}$ denote the OAM phase distribution of the $n_t$th and $n_r$th modes, respectively, ``${\rm rot}\left(\cdot,[\phi_{x},\phi_{y},\phi_{z}]\right)$'' represents the rotation of $\phi_{x}$, $\phi_{y}$, and $\phi_{z}$ rads around the $x$-, $y$-, and $z$- axes according to the right-hand spiral law, $[\phi^{\rm Tx}_{x},\phi^{\rm Tx}_{y},\phi^{\rm Tx}_{z}]$ and $[\phi^{\rm Rx}_{x},\phi^{\rm Rx}_{y},\phi^{\rm Rx}_{z}]$ denote the rotation angles for the transmit and receive OAM antennas, respectively. In Eq.~\eqref{eq:OAM_phase}, $\hat{a}_{n_t}$ and $\hat{a}_{n_r}$ can be expressed as follows:
\begin{align}
&\hat{a}_{m}\left(\theta,\varphi\right)=
\begin{cases}
e^{jm\varphi},\ {\rm for}\ \theta\in[0,\pi/2];\\
\hspace{0.27cm}0\hspace{0.27cm},\ {\rm for\ otherwise},
\end{cases}
\end{align}
where $m$ denotes the corresponding OAM mode. For transmit UCA antenna with $N_t$ elements and receive UCA antenna with $N_r$ elements, the $n_t$th transmit OAM mode and the $n_r$th receive OAM mode, denoted by $m_{n_t}$ and $m_{n_r}$ respectively, can be given as follows:
\begin{align}
\begin{cases}
m_{n_t} = \lfloor\frac{2-Nt}{2}\rfloor+n_t-1;\\
m_{n_r} = \lfloor\frac{2-Nr}{2}\rfloor+n_r-1.
\end{cases}
\end{align}

The discrete form of $H_{n_r,n_t}(f)$ can be expressed as follows:
%\begin{align}
%H_{n_r,n_t}(k)=\sum_{l=1}^L&\alpha_lc_{r,n_r}\left(\boldsymbol{\mathrm \Omega}_{r,l}\right)c^*_{t,n_t}\left(\boldsymbol{\mathrm \Omega}_{t,l}\right)\exp{\left[ja_{t,n_t}\left(\theta_{t,l},\varphi_{t,l}\right)-ja_{r,n_r}\left(\theta_{r,l},\varphi_{r,l}\right)\right]}\exp{\left(-j2\pi f_k\tau_l\right)}=\sum_{l=1}^Ls_{n_r,n_t}\left(\boldsymbol{\mathrm \Theta}_l,k\right),
%\label{eq:CTF_d}
%\end{align}
\begin{align}
H_{n_r,n_t}(k)=&\sum_{l=1}^L\alpha_lc_{r,n_r}\left(\boldsymbol{\mathrm \Omega}_{r,l}\right)c^*_{t,n_t}\left(\boldsymbol{\mathrm \Omega}_{t,l}\right)\nonumber\\
&\quad\exp{\left[ja_{t,n_t}\left(\theta_{t,l},\varphi_{t,l}\right)-ja_{r,n_r}\left(\theta_{r,l},\varphi_{r,l}\right)\right]}\nonumber\\
&\quad\quad\exp{\left(-j2\pi f_k\tau_l\right)}=\sum_{l=1}^Ls_{n_r,n_t}\left(\boldsymbol{\mathrm \Theta}_l,k\right),
\label{eq:CTF_d}
\end{align}
where $f_k$ denotes the frequency at the $k$th sampled frequency point with $1\le f_k\le N_f$, $s_{n_r,n_t}\left(\boldsymbol{\mathrm \Theta}_l,k\right)$ represents the CTF of the $l$th MPC at $f_k$, and $\boldsymbol{\mathrm \Theta}_l=\left[\alpha_l,\tau_l,\theta_{t,l},\varphi_{t,l},\theta_{r,l},\varphi_{r,l}\right]$ denotes the parameters of the $l$th MPC. Thus, the matrix form of the CTF, denoted by $\boldsymbol{\mathrm H}\in \mathbb{C}^{N_r\times N_t\times N_f}$, can be given as follows:
%\begin{align}
%\boldsymbol{\mathrm H}=\sum_{l=1}^L&\alpha_l\left[\boldsymbol{\mathrm c}_{r}\left(\boldsymbol{\mathrm \Omega}_{r,l}\right)\cdot e^{-j\boldsymbol{\mathrm a}_{r}\left(\theta_{r,l},\varphi_{r,l}\right)}\right]\left[\boldsymbol{\mathrm c}_{t}\left(\boldsymbol{\mathrm \Omega}_{t,l}\right)\cdot e^{-j\boldsymbol{\mathrm a}_{t}\left(\theta_{t,l},\varphi_{t,l}\right)}\right]^H\otimes\exp{\left(-j2\pi \boldsymbol{\mathrm f}\tau_l\right)}
%=\sum_{l=1}^L\boldsymbol{\mathrm s}\left(\boldsymbol{\mathrm \Theta}_l\right),
%\label{eq:CTF_H}
%\end{align}
\begin{align}
\boldsymbol{\mathrm H}\hspace{-0.1cm}&=\hspace{-0.1cm}\sum_{l=1}^L\hspace{-0.1cm}\alpha_l\hspace{-0.1cm}\left[\boldsymbol{\mathrm c}_{r}\left(\boldsymbol{\mathrm \Omega}_{r,l}\right)\hspace{-0.05cm}\cdot\hspace{-0.05cm} e^{-j\boldsymbol{\mathrm a}_{r}\left(\theta_{r,l},\varphi_{r,l}\right)}\right]\hspace{-0.1cm}\left[\boldsymbol{\mathrm c}_{t}\left(\boldsymbol{\mathrm \Omega}_{t,l}\right)\hspace{-0.05cm}\cdot\hspace{-0.05cm} e^{-j\boldsymbol{\mathrm a}_{t}\left(\theta_{t,l},\varphi_{t,l}\right)}\right]^H\nonumber\\
&\quad\quad\quad\otimes\exp{\left(-j2\pi \boldsymbol{\mathrm f}\tau_l\right)}\nonumber\\
\hspace{-0.1cm}&\quad=\hspace{-0.1cm}\sum_{l=1}^L\boldsymbol{\mathrm s}\left(\boldsymbol{\mathrm \Theta}_l\right),
\label{eq:CTF_H}
\end{align}
where $\boldsymbol{\mathrm f}=\left[f_1,f_2,\cdots,f_{N_f}\right]$ denotes the $N_f$ sampled frequency points, $\boldsymbol{\mathrm s}\left(\boldsymbol{\mathrm \Theta}_l\right)$ denotes the CTF of the $l$th MPC, $\boldsymbol{\mathrm c}_{r}\left(\boldsymbol{\mathrm \Omega}_{r,l}\right)\in \mathbb{C}^{N_r\times 1}$ and $\boldsymbol{\mathrm c}_{t}\left(\boldsymbol{\mathrm \Omega}_{t,l}\right)\in \mathbb{C}^{N_t\times 1}$ denotes the array antenna patterns of the receive antenna array and the transmit antenna array, respectively. Besides, $\boldsymbol{\mathrm a}_{r}\left(\theta_{r,l},\varphi_{r,l}\right)\in \mathbb{C}^{N_r\times 1}$ and $\boldsymbol{\mathrm a}_{t}\left(\theta_{t,l},\varphi_{t,l}\right)\in \mathbb{C}^{N_t\times 1}$ represents the phases of the $l$th MPC that attribute to the $N_r$ receive OAM-modes and the $N_t$ transmit OAM-modes, respectively.

\begin{table*}[ht]
\centering
\renewcommand{\arraystretch}{1.3} % Default value: 1
\begin{tabular}{|c|c|}
\hline
Scenario & Path Loss [dB]  \\
\hline
Indoor LOS & $8.692+ 16.69\lg\left(\left|J_{m_{n_t}}\left(262.9f^{\rm GHz}\frac{R_r}{\sqrt{R_r^2 + d^2}}\right)\right|\left|J_{m_{n_r}}\left(262.9f^{\rm GHz}\frac{R_t}{\sqrt{R_t^2 + d^2}}\right)\right|\right) + 17.3\lg(d) + 20\lg(f^{\rm GHz})$ \\
\hline
Through-wall & $18.58+ 6.064\lg\left(\left|J_{m_{n_t}}\left(47.65f^{\rm GHz}\frac{R_r}{\sqrt{R_r^2 + d^2}}\right)\right|\left|J_{m_{n_r}}\left(47.65f^{\rm GHz}\frac{R_t}{\sqrt{R_t^2 + d^2}}\right)\right|\right) + 17.3\lg(d) + 24.9\lg(f^{\rm GHz})$ \\
\hline
Outdoor LOS& $18.65+ 10.16\lg\left(\left|J_{m_{n_t}}\left(418.1f^{\rm GHz}\frac{R_r}{\sqrt{R_r^2 + d^2}}\right)\right|\left|J_{m_{n_r}}\left(418.1f^{\rm GHz}\frac{R_t}{\sqrt{R_t^2 + d^2}}\right)\right|\right) + 17.3\lg(d) + 20\lg(f^{\rm GHz})$ \\
\hline
\end{tabular}
\vspace{5pt}
\caption{Path loss of the same OAM mode.}\label{tab:pathloss}
\end{table*}
\section{Simulation Channel Generation}\label{sec:channelGen}
\begin{figure*}[htbp]
\centering
%\vspace{-10pt}
\includegraphics[scale=0.53]{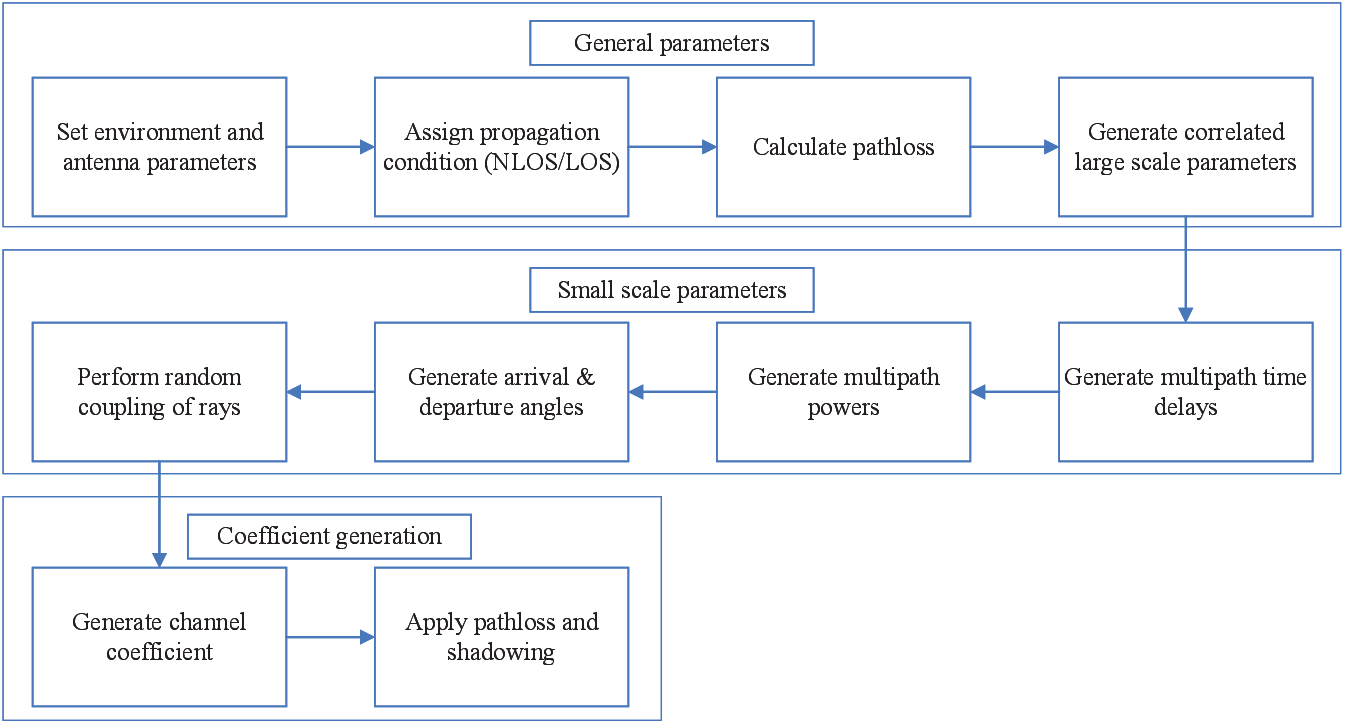}
%\vspace{-15pt}
\caption{Channel coefficient generation procedure.} \label{fig:flowChart}
%\vspace{-5pt}
\end{figure*}
Based on the channel model and the analyses of measured data, we can give the steps to generate the simulation OAM channel as in Fig.~\ref{fig:flowChart} in page 8. Detailed steps are as follows:

1) Set environment and antenna parameters: Define the simulation scenario and establish a global Cartesian coordinate system along with the corresponding spherical coordinate system. Specify the number of modes for both transmitting and receiving antennas. Based on the global coordinate system, establish the positions and orientations of the transmitting and receiving antennas. Calculate LOS departure and arrival angles between the transmitting and receiving antennas. Present the complex direction pattern for each mode of transmitting and receiving antennas based on their respective directions. Compute the departure and arrival directions based on the antenna orientations. Lastly, determine the center frequency and bandwidth.

2) Calculate the path loss: Calculate the path loss based on the path loss formulas fitted by the measurement results. The path loss calculation formulas for the same-mode transmission scenario are given as in Table~\ref{tab:pathloss}.
%\begin{table*}[ht]
%\centering
%\renewcommand{\arraystretch}{1.3} % Default value: 1
%\begin{tabular}{|c|c|}
%\hline
%Scenario & Path Loss [dB]  \\
%\hline
%Indoor LOS & $8.692+ 16.69\lg\left(\left|J_{m_{n_t}}\left(262.9f^{\rm GHz}\frac{R_r}{\sqrt{R_r^2 + d^2}}\right)\right|\left|J_{m_{n_r}}\left(262.9f^{\rm GHz}\frac{R_t}{\sqrt{R_t^2 + d^2}}\right)\right|\right) + 17.3\lg(d) + 20\lg(f^{\rm GHz})$ \\
%\hline
%Through-wall & $18.58+ 6.064\lg\left(\left|J_{m_{n_t}}\left(47.65f^{\rm GHz}\frac{R_r}{\sqrt{R_r^2 + d^2}}\right)\right|\left|J_{m_{n_r}}\left(47.65f^{\rm GHz}\frac{R_t}{\sqrt{R_t^2 + d^2}}\right)\right|\right) + 17.3\lg(d) + 24.9\lg(f^{\rm GHz})$ \\
%\hline
%Outdoor LOS& $18.65+ 10.16\lg\left(\left|J_{m_{n_t}}\left(418.1f^{\rm GHz}\frac{R_r}{\sqrt{R_r^2 + d^2}}\right)\right|\left|J_{m_{n_r}}\left(418.1f^{\rm GHz}\frac{R_t}{\sqrt{R_t^2 + d^2}}\right)\right|\right) + 17.3\lg(d) + 20\lg(f^{\rm GHz})$ \\
%\hline
%\end{tabular}
%\caption{Path loss of the same OAM mode.}\label{tab:pathloss}
%\end{table*}

3) Generate large scale parameters: Based on the scenario, chose corresponding large scale parameters, including the RMS DS, RMS AS, shadow fading standard deviation and other parameters.

4) Generate multipath time delays: According to the scenario and measurement results, multipath delay is generated by RMS DS. The delay of the $l$th path is given as $\tau_l = -r_{\tau}\tau_{RMS}\ln(X_l)$, where $r_{\tau}=1.5$ is the delay distribution proportionality factor and $X_l$ denotes a $(0,1)$ uniform distribution. Then, normalise the delays by subtracting the minimum delay and sort the normalised delays to ascending order as $\tau_l = {\rm sort}(\tau_l-{\rm min}(\tau_l))$.

In LOS scenario, additional scaling is required to compensate for the effect of LOS peak addition to the DS as $\tau_l = \tau_l/(0.7705-0.0433K+0.0002K^2+0.000017K^3)$, where $K$ denotes the Ricean K-factor given in Step 2.

5) Generate multipath powers: Calculate multipath powers from multipath time delays and RMS DS. The normalized multipath power of the $l$th path, denoted by $P_l$, can be given as $P_l = \frac{\exp\left(\tau_l\frac{r_{\tau}-1}{r_{\tau}\tau_{\rm RMS}}\right)10^{-Z_l/10}}{\sum_l\exp\left(\tau_l\frac{r_{\tau}-1}{r_{\tau}\tau_{\rm RMS}}\right)10^{-Z_l/10}}$, where $Zn\sim N(0,\zeta^2)$ is the per cluster shadowing term in dB and $\zeta$ denotes the per cluster shadowing standard deviation. It should be noticed that $\tau_l$ here used to generate multipath powers in LOS scenario is not scaled as in Step 3.

In LOS scenario, and additional power component is added to the first cluster as $P_l = \frac{P_l}{K_R+1} + \delta(l-1)\frac{K_R}{K_R+1}$, where $K_R$ denotes Ricean K-factor given in Step 2 converted to linear scale. Then the multipath powers of each ray in the cluster can be given as $P_l/M$, where $M$ is the number of rays per cluster given in Step 2.

\begin{figure*}[bp]
\centering
\vspace{-10pt}
\includegraphics[scale=0.27]{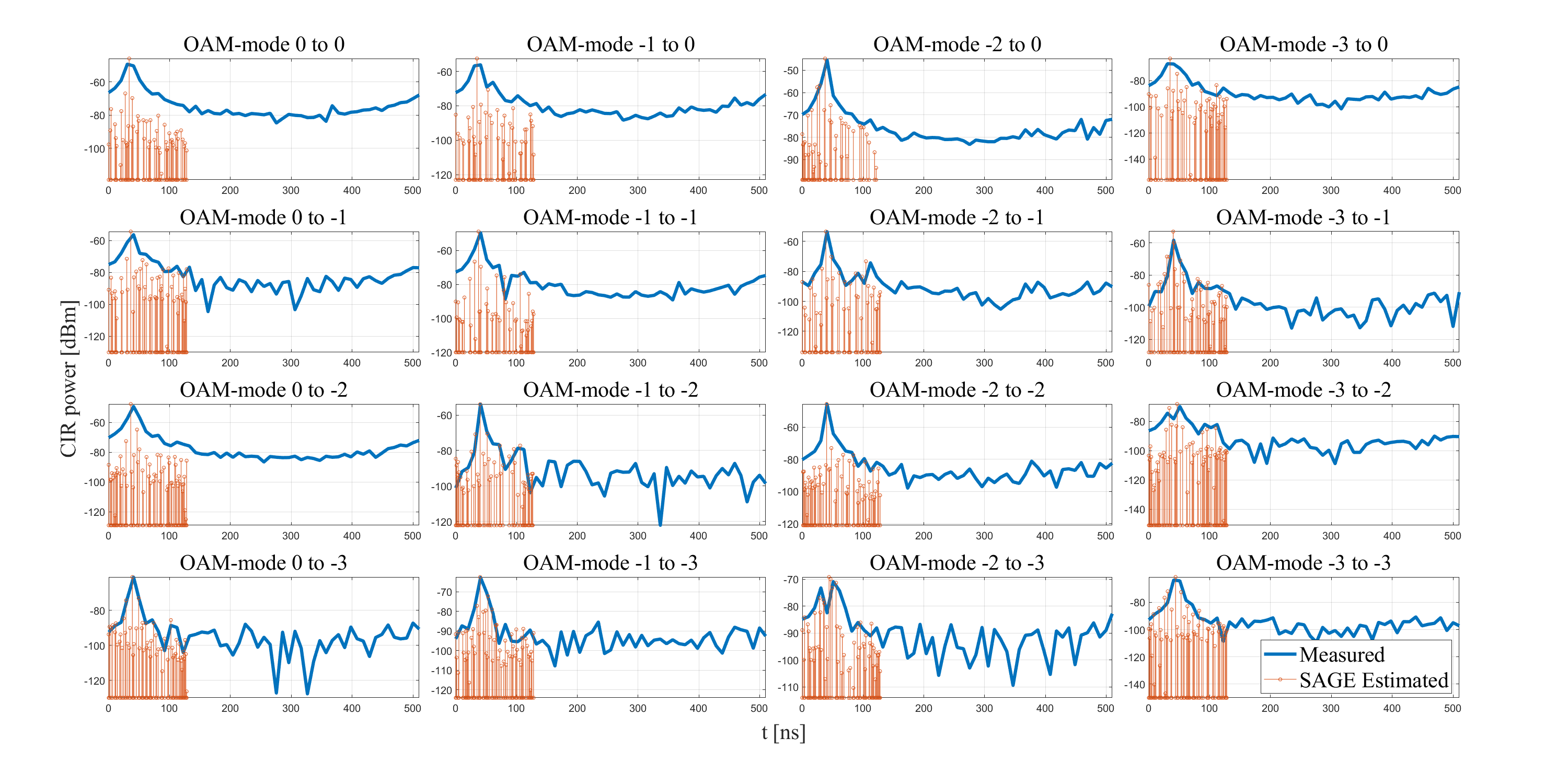}
\vspace{-15pt}
\caption{SAGE estimated delay PSD VS. measured CIR.} \label{fig:delayPSD}
\vspace{-5pt}
\end{figure*}

6) Generate angle of arrival and angle of departure for azimuth angle and elevation angle parameters respectively: Data is expanded from multipath RMS AS to generate random arrival and departure azimuths and elevation angles. The formula for generating AOAs is as follows:
\begin{align}
\varphi_{r,l} \hspace{-0.1cm}=\hspace{-0.1cm}
\begin{cases}
\hspace{-0.1cm}\left(\frac{2(\varphi_{r,\rm RMS}/1.4)\sqrt{-\ln(P_l/\max(P_l))}}{C_\varphi}X_l\hspace{-0.1cm}+\hspace{-0.1cm}Y_l\right)\\
\hspace{-0.1cm}\quad\hspace{-0.1cm}-\hspace{-0.1cm}\left(\hspace{-0.1cm}\frac{2(\varphi_{r,\rm RMS}/1.4)\sqrt{-\ln(P_1/\max(P_1))}}{C_\varphi}X_1\hspace{-0.1cm}+\hspace{-0.1cm}Y_1\hspace{-0.1cm}-\hspace{-0.1cm}\varphi_{r,\rm LOS}\hspace{-0.1cm}\right)\hspace{-0.1cm},\\
\hspace{6cm}{\rm for\ LOS};\\
\hspace{-0.1cm}\frac{2(\varphi_{r,\rm RMS}/1.4)\sqrt{-\ln(P_l/\max(P_l))}}{C_\varphi}X_l\hspace{-0.1cm}+\hspace{-0.1cm}Y_l\hspace{-0.1cm}+\hspace{-0.1cm}\varphi_{r,\rm LOS},\\
\hspace{6cm}{\rm for\ NLOS},
\end{cases}
\end{align}
where $Y_l\sim N(0,(\varphi_{r,\rm LOS}/7)^2)$ is a Gaussian function, $\varphi_{r,\rm LOS}$ denotes the azimuth direction of arrival corresponding to the transmit antenna given in Step 1, and $C_\varphi$ is the scaling factor defined as follows:
\begin{align}
C_\varphi=
\begin{cases}
C_\varphi^{\rm NLOS}(1.1035\hspace{-0.1cm}-\hspace{-0.1cm}0.028K\hspace{-0.1cm}-\hspace{-0.1cm}0.002K^2\hspace{-0.1cm}+\hspace{-0.1cm}0.0001K^3),\\
\hspace{6cm}{\rm for\ LOS};\\
C_\varphi^{\rm NLOS},\ {\rm for\ NLOS},
\end{cases}
\end{align}
where $C_\varphi^{\rm NLOS}$ is defined as a scaling factor related to the total number of clusters given as $C_\varphi^{\rm NLOS}=15$ in this paper.

To further generate AOA of each ray, denoted by $\varphi_{r,l,m}$, an angle offset is added to the cluster angles as $\varphi_{r,l,m} = \varphi_{r,l} + C_{ASA}\alpha_m$, where $C_{ASA}$ denotes the cluster-wise RMS azimuth spread of arrival angles as given in Step 2 and $\alpha_m$ is the ray offset angles within a cluster. AODs are generated in a similar method.

 %given as in Table~\ref{tab:rayOffset}.
%\begin{table}[ht]
%\centering
%\renewcommand{\arraystretch}{1.3} % Default value: 1
%\begin{tabular}{|c|c|}
%\hline
%Ray number $m$ & Basis vector of offset angles $\alpha_m$  \\
%\hline
%1,2 & $\pm 0.0447$ \\
%\hline
%3,4 & $\pm 0.1413$ \\
%\hline
%5,6 & $\pm 0.2492$ \\
%\hline
%7,8 & $\pm 0.3715$ \\
%\hline
%9,10 & $\pm 0.5129$ \\
%\hline
%11,12 & $\pm 0.6769$ \\
%\hline
%13,14 & $\pm 0.8844$ \\
%\hline
%15,16 & $\pm 1.1481$ \\
%\hline
%17,18 & $\pm 1.5195$ \\
%\hline
%19,20 & $\pm 2.1551$ \\
%\hline
%\end{tabular}
%\caption{Ray offset angles within a cluster, given for RMS angle spread normalized to 1.}\label{tab:rayOffset}
%\end{table}

EOAs are given as follows:
\begin{align}
\theta_{r,l} \hspace{-0.1cm}=\hspace{-0.1cm}
\begin{cases}
\frac{\theta_{r,\rm RMS}\ln(P_l/\max(P_l))}{C_\theta}X_l-Y_l\\
\quad+\hspace{-0.1cm}\left(-\frac{\theta_{r,\rm RMS}\ln(P_1/\max(P_1))}{C_\theta}X_1\hspace{-0.1cm}+\hspace{-0.1cm}Y_1\hspace{-0.1cm}+\hspace{-0.1cm}\theta_{r,\rm LOS}\right),\\
\hspace{5.7cm}{\rm for\ LOS},\\
\frac{\theta_{r,\rm RMS}\ln(P_l/\max(P_l))}{C_\theta}X_l-Y_l+\theta_{r,\rm LOS},\ {\rm for\ NLOS},
\end{cases}
\end{align}
where $\theta_{r,\rm LOS}$ denotes the elevation direction of arrival corresponding to the transmit antenna as given in Step 2. To further generate EOA of each ray, denoted by $\theta_{r,l,m}$, an angle offset is added to the cluster angles as $\theta_{r,l,m} = \theta_{r,l} + C_{ESA}\alpha_m$, where $C_{ESA}$ denotes the cluster-wise RMS azimuth spread of arrival angles as given in Step 2.

EODs are given in a similar way as generation of EOAs as follows:
\begin{align}
\theta_{t,l} \hspace{-0.1cm}=\hspace{-0.1cm}
\begin{cases}
\frac{\theta_{t,\rm RMS}\ln(P_l/\max(P_l))}{C_\theta}X_l-Y_l\\
\quad+\left(-\frac{\theta_{t,\rm RMS}\ln(P_1/\max(P_1))}{C_\theta}X_1+Y_1+\theta_{t,\rm LOS}\right),\\
\hspace{6cm}{\rm for\ LOS},\\
\frac{\theta_{t,\rm RMS}\ln(P_l/\max(P_l))}{C_\theta}X_l-Y_l+\theta_{t,\rm LOS}-\mu_{\rm offset,EOD},\\
\hspace{6cm}{\rm for\ NLOS},
\end{cases}
\end{align}
where $\theta_{t,\rm LOS}$ denotes the elevation direction of departure corresponding to the receive antenna and $\mu_{\rm offset,EOD}$ denotes the EOD mean offset as given in Step 2. To further generate EOD of each ray, denoted by $\theta_{t,l,m}$, an angle offset is added to the cluster angles as $\theta_{t,l,m} = \theta_{t,l} + (3/8)10^{\mu_{\rm lgEOD}}\alpha_m$, where $\mu_{\rm lgEOD}$ denotes the mean value ESD log-normal distribution.

7) Perform random coupling of rays: The AOA and AOD are combined into twins randomly.

8) Generate channel coefficient: Substitute the generated multipath parameters into the channel model CTF expression to generate a normalized CTF matrix as follows:
\begin{align}
H_{n_r,n_t}(f)=&\sum_{l=1}^L\alpha_lc_{r,n_r}\left(\boldsymbol{\mathrm \Omega}_{r,l}\right)c^*_{t,n_t}\left(\boldsymbol{\mathrm \Omega}_{t,l}\right)\nonumber\\
&\quad\exp{\left[ja_{t,n_t}\left(\theta_{t,l},\varphi_{t,l}\right)-ja_{r,n_r}\left(\theta_{r,l},\varphi_{r,l}\right)\right]}\nonumber\\
&\quad\quad\exp{\left(-j2\pi f\tau_l\right)},
\end{align}

9) Apply pathloss and shadowing: The normalized CTF is multiplied by the path loss and shadow fading in the corresponding scene to generate the final CTF, denoted by $\hat{\boldsymbol{\mathrm H}}_{n_r,n_t}$. The combined path loss and shadow fading from the $n_t$th transmit OAM mode to the $n_r$th receive OAM mode at frequency $f$, denoted by $\hat{H}_{n_r,n_t}(f)$, is given as follows:
\begin{align}
\hat{H}_{n_r,n_t}(f)\hspace{-0.1cm}=&PL_{n_r,n_t}(d)\sum_{l=1}^L\alpha_lc_{r,n_r}\left(\boldsymbol{\mathrm \Omega}_{r,l}\right)c^*_{t,n_t}\left(\boldsymbol{\mathrm \Omega}_{t,l}\right)\nonumber\\
&\quad\exp{\left[ja_{t,n_t}\left(\theta_{t,l},\varphi_{t,l}\right)-ja_{r,n_r}\left(\theta_{r,l},\varphi_{r,l}\right)\right]}\nonumber\\
&\quad\quad\exp{\left(-j2\pi f\tau_l\right)},
\end{align}
where $PL_{n_r,n_t}(d)=10^{PL^{\rm dB}_{n_r,n_t}(d)/10}$ denotes the path loss from the $n_t$th transmit OAM mode to the $n_r$th received OAM mode.

\section{Channel Measurement and Simulation Results}\label{sec:results}
%\section{Channel Measurement Results Analyses}\label{sec:results}
%In this part, we give the channel measurement results for verifying our channel model as well as providing some further insights. The measurement results are given based on analyses of measured CTFs.
In this part, we give the channel measurement and simulation results for verifying our channel model as well as providing some further insights. The measurement results are given based on analyses of measured CTFs. The simulation results are given based on our proposed OAM channel model and parameters estimated from the measurement data, including RMSs of delay, angles of arrival, angles of departure.

\subsection{Delay PSD}
%\begin{figure*}[htbp]
%\centering
%\vspace{-10pt}
%\includegraphics[scale=0.27]{pics//delayPSD_SageEtracted.eps}
%\vspace{-15pt}
%\caption{SAGE estimated delay PSD VS. measured CIR.} \label{fig:delayPSD}
%%\vspace{-5pt}
%\end{figure*}
Figure~\ref{fig:delayPSD} shows a set of measured CIR and SAGE-estimated delay PSD for an indoor LOS scenario at $5.8$ GHz with a transmitter-to-receiver distance of $9.6$ meters, as an example. The MPC number was set to $100$. Figure~\ref{fig:delayPSD} demonstrates that the estimated delay PSD aligns well with the measured results. Additionally, Fig.~\ref{fig:delayPSD} indicates significant crosstalk between different OAM modes, even in the LOS scenario, making it challenging to separate signals of different OAM modes.

\begin{figure}[htbp]
\centering
%\vspace{-10pt}
\includegraphics[scale=0.45]{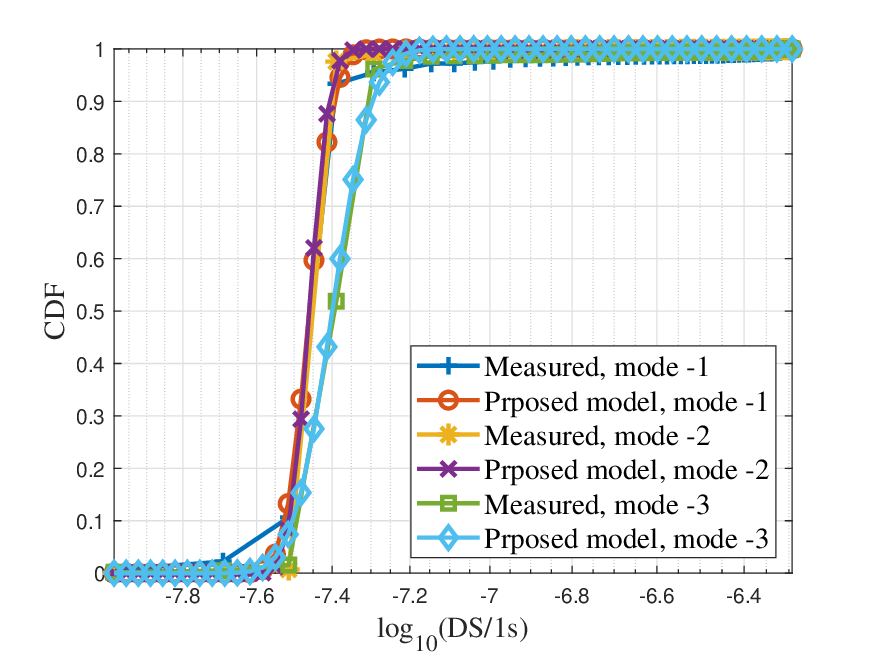}
\vspace{-5pt}
\caption{CDFs of ${\rm log}_{10}(\rm DS/1\ \rm s)$ for OAM mode $-3$ to $-1$ in indoor LOS scenario at $5.8$ GHz.} \label{fig:delayCDF}
%\vspace{-5pt}
\end{figure}
%Figure~\ref{fig:delayCDF} shows the measured and simulated cumulative distribution functions (CDFs) of ${\rm log}_{10}(\rm DS/1\ \rm s)$ for OAM mode $-3$ to $-1$ in indoor LOS scenario at $5.8$ GHz with a transmitter-to-receiver distance of $9.6$ meters. The simulated CDFs are plotted based on the normal fitting parameters of the SAGE-estimated delay PSD. Figure~\ref{fig:delayCDF} shows that the proposed general channel model are consistent with the measurement data, verifying our channel model's accuracy. We can also find in Fig.~\ref{fig:delayCDF} that the DS CDFs of OAM mode $-1$ and mode $-2$ are very close. However, the mean value of DS of OAM mode $-3$ is lager than those of the other two modes. DS of OAM mode $-3$ also has a larger variance.

Figure~\ref{fig:delayCDF} shows the measured and simulated cumulative distribution functions (CDFs) of $\log_{10}(\text{DS}/1\ \text{s})$ for OAM modes $-3$ to $-1$ in an indoor LOS scenario at $5.8$ GHz with a transmitter-to-receiver distance of $9.6$ meters. The simulated CDFs were plotted based on the normal fitting parameters of the SAGE-estimated delay PSD. Figure~\ref{fig:delayCDF} demonstrates that the proposed general channel model is consistent with the measurement data, verifying the accuracy of our channel model. Additionally, Fig.~\ref{fig:delayCDF} reveals that the DS CDFs of OAM modes $-1$ and $-2$ are very close. However, the mean value of DS for OAM mode $-3$ is larger than those of the other two modes, and DS for OAM mode $-3$ also has a larger variance.

\subsection{Angular PSD}
\begin{figure}[htbp]
\centering
%\vspace{-15pt}
\subfigure[AoA PSD.]{
\begin{minipage}{0.45\linewidth}
\centering
\includegraphics[scale=0.27]{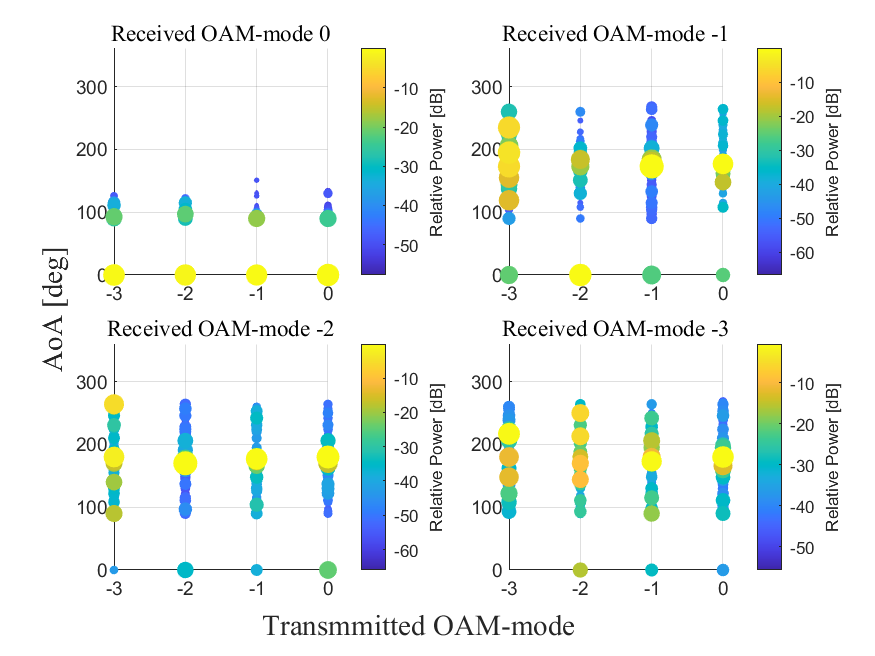}\label{fig:PSD_indoor_AoA}
%\vspace{-5pt}
\end{minipage}
}
\subfigure[EoA PSD.]{
\begin{minipage}{0.45\linewidth}
\centering
\includegraphics[scale=0.27]{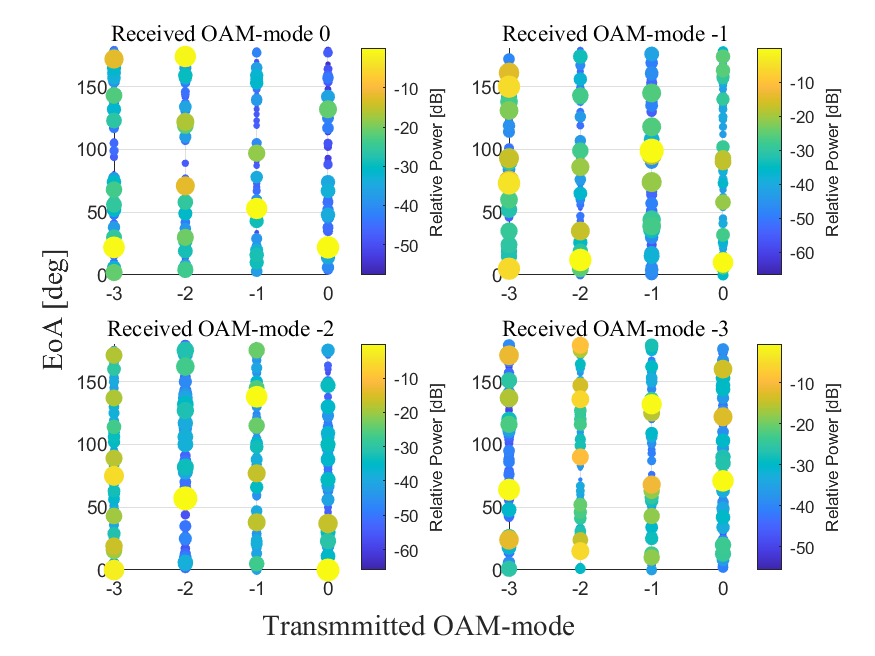}\label{fig:PSD_indoor_EoA}
%\vspace{-5pt}
\end{minipage}
}
\subfigure[AoD PSD.]{
\begin{minipage}{0.45\linewidth}
\centering
\includegraphics[scale=0.27]{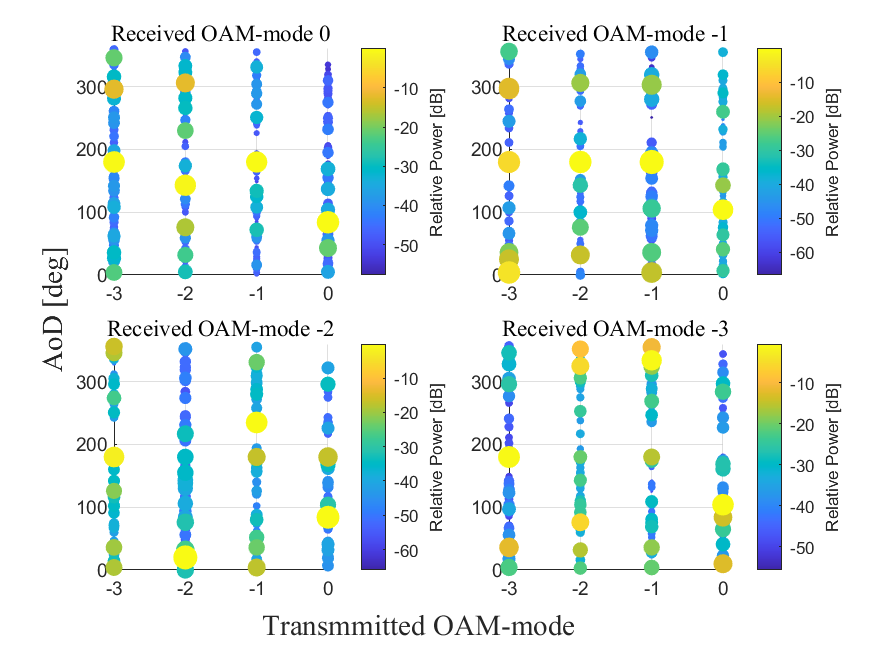}\label{fig:PSD_indoor_AoD}
%\vspace{-5pt}
\end{minipage}
}
\subfigure[EoD PSD.]{
\begin{minipage}{0.45\linewidth}
\centering
\includegraphics[scale=0.27]{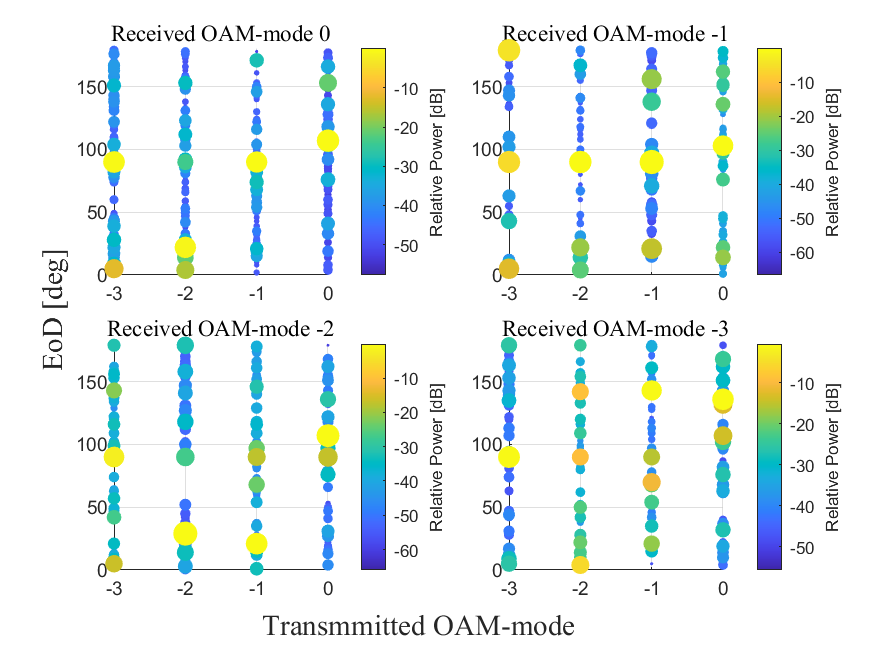}\label{fig:PSD_indoor_EoD}
%\vspace{-5pt}
\end{minipage}
}
\centering
\vspace{-10pt}
\caption{SAGE estimated angular PSDs at $5.8$ GHz in indoor LOS scenario.}\label{fig:AngularPSD}
%\vspace{-10pt}
\end{figure}
\begin{figure}[htbp]
\centering
%\vspace{-15pt}
\subfigure[AoA PSD.]{
\begin{minipage}{0.45\linewidth}
\centering
\includegraphics[scale=0.27]{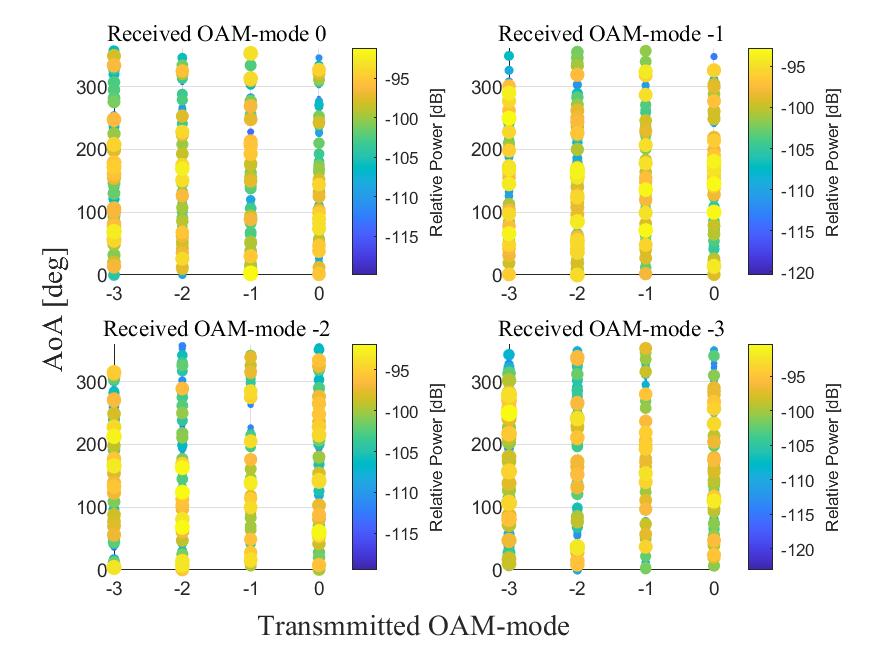}\label{fig:PSD_indoorLoS_28G_AoA}
%\vspace{-5pt}
\end{minipage}
}
\subfigure[EoA PSD.]{
\begin{minipage}{0.45\linewidth}
\centering
\includegraphics[scale=0.27]{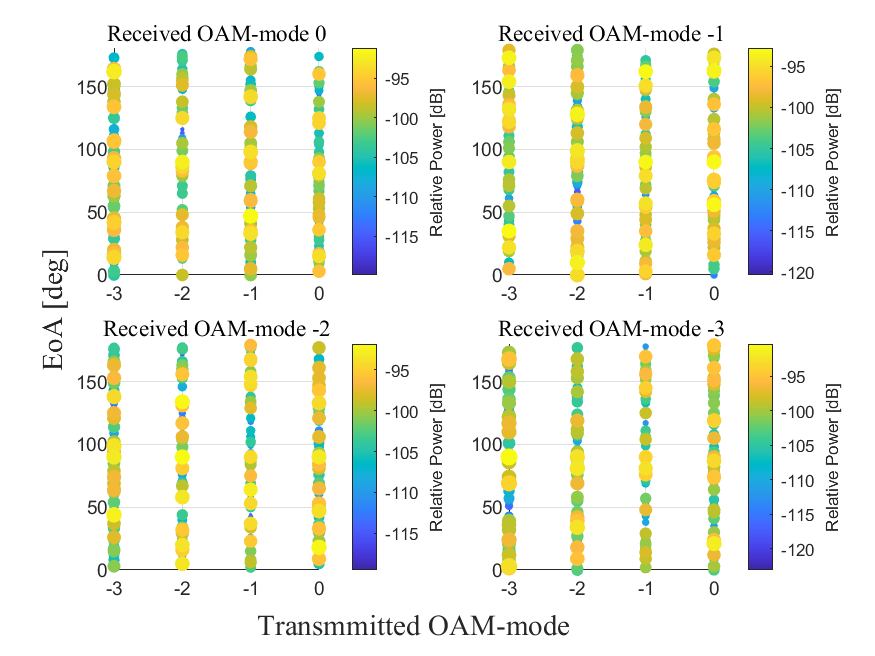}\label{fig:PSD_indoorLoS_28G_EoA}
%\vspace{-5pt}
\end{minipage}
}\\
\subfigure[AoD PSD.]{
\begin{minipage}{0.45\linewidth}
\centering
\includegraphics[scale=0.27]{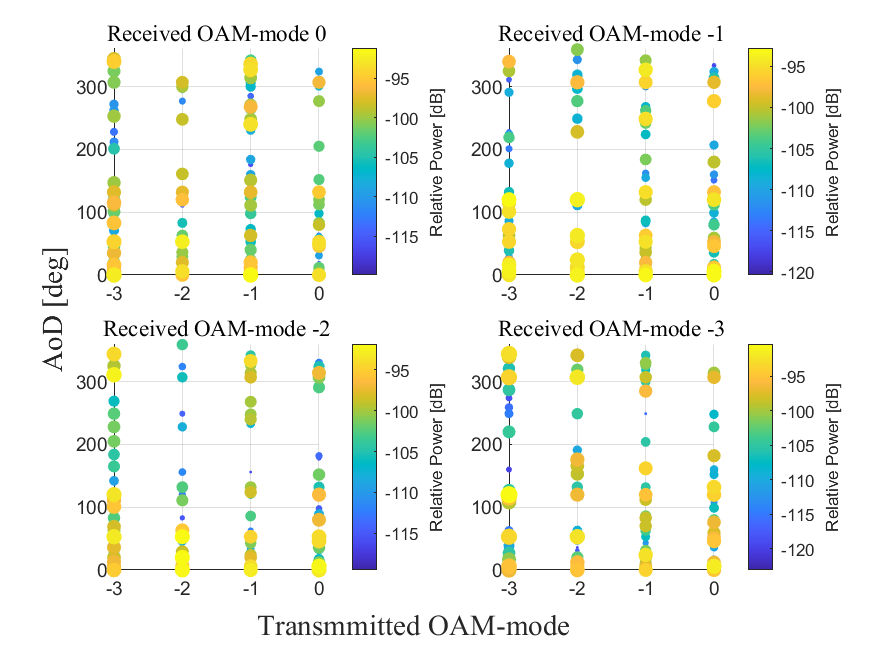}\label{fig:PSD_indoorLoS_28G_AoD}
%\vspace{-5pt}
\end{minipage}
}
\subfigure[EoD PSD.]{
\begin{minipage}{0.45\linewidth}
\centering
\includegraphics[scale=0.27]{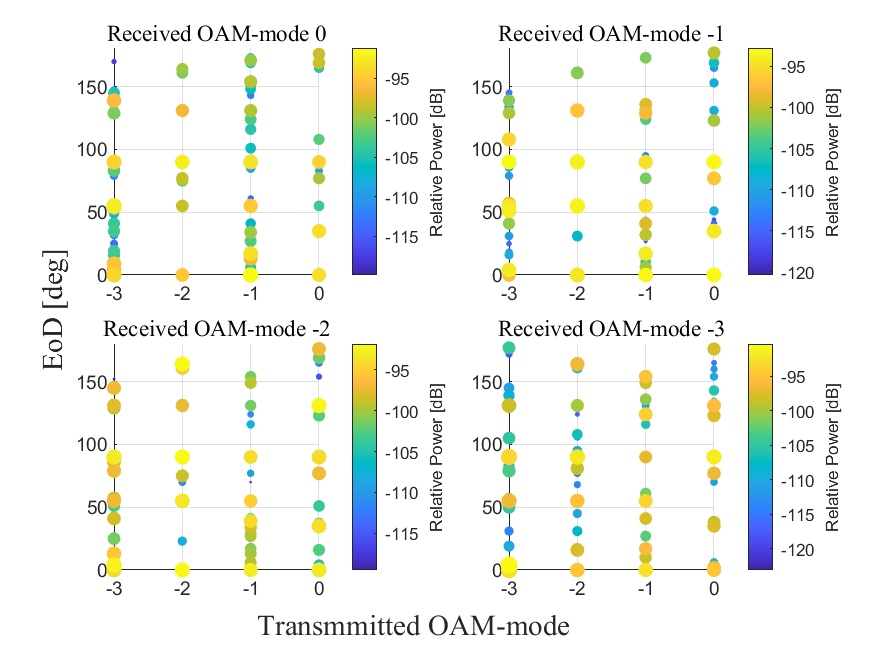}\label{fig:PSD_indoorLoS_28G_EoD}
%\vspace{-5pt}
\end{minipage}
}
\centering
\vspace{-10pt}
\caption{SAGE estimated angular PSDs at $28$ GHz in indoor LOS scenario.}\label{fig:AngularPSD28G}
%\vspace{-10pt}
\end{figure}
%Figures~\ref{fig:AngularPSD} and \ref{fig:AngularPSD28G} show the SAGE estimated angular PSDs of OAM modes $-3$ to $0$, where Fig.~\ref{fig:AngularPSD} gives a set of SAGE estimated angular PSDs for indoor LOS scenario at $5.8$ GHz with the transmitter-to-receiver distance of $9.6$ m and Fig.~\ref{fig:AngularPSD28G} shows a set of SAGE-estimated angular PSDs for an indoor LOS scenario at $28$ GHz with a transmitter-to-receiver distance of $10$ meters. We use the size and color of MPC points to represent the normalized relative powers of MPCs in dB. Figure~\ref{fig:AngularPSD} indicates that the extracted MPC powers remain relatively concentrated in the LOS scenario at $5.8$ GHz. However, the direction of the power is not completely aligned with the antenna direction. Additionally, it is clear that the multipath effect is significantly more severe in the $28$ GHz band than in the $5.8$ GHz band when comparing Fig~\ref{fig:AngularPSD} with \ref{fig:AngularPSD28G}. The normalized relative powers are much lower than those at $5.8$ GHz.

Figures~\ref{fig:AngularPSD} and \ref{fig:AngularPSD28G} present the SAGE-estimated angular PSDs for OAM modes $-3$ to $0$ of indoor LOS scenarios at distinct frequencies. In these figure, the size and color of the MPC points are used to represent the normalized relative powers of the MPCs in dB. Figure~\ref{fig:AngularPSD} illustrates a set of SAGE-estimated angular PSDs for an indoor LOS scenario at $5.8$ GHz, where the distance between the transmitter and receiver is $9.6$ meters. The data indicate that the extracted MPC powers are relatively concentrated. However, it is noteworthy that the direction of these powers does not align perfectly with the antenna direction, which may suggest minor discrepancies due to OAM divergence and environmental reflections. Figure~\ref{fig:AngularPSD28G} shows a set of SAGE-estimated angular PSDs for an indoor LOS scenario at $28$ GHz, with the transmitter-to-receiver distance set at $10$ meters. Comparing Fig.~\ref{fig:AngularPSD28G} to Fig.~\ref{fig:AngularPSD}, it is clear that the multipath effect is significantly more severe at $28$ GHz than at $5.8$ GHz. This severity is evidenced by the more dispersed and lower normalized relative powers at $28$ GHz.

%The pronounced difference in multipath effects between the two frequencies can be attributed to the inherent properties of higher-frequency signals. At $28$ GHz, signals are more susceptible to attenuation and scattering, resulting in a greater reduction in power and a more complex multipath environment. This complexity can lead to lower relative power levels and more significant variations in signal direction, as observed in the figures. These findings underscore the challenges associated with higher-frequency signal propagation in indoor environments, where obstacles and reflective surfaces can drastically influence the signal behavior.

%\begin{figure}[htbp]
%\centering
%%\vspace{-15pt}
%\subfigure[AoA PSD.]{
%\begin{minipage}{0.45\linewidth}
%\centering
%\includegraphics[scale=0.3]{pics//PSD_outdoor_AoA.eps}\label{fig:PSD_outdoor_AoA}
%%\vspace{-5pt}
%\end{minipage}
%}
%\subfigure[EoA PSD.]{
%\begin{minipage}{0.45\linewidth}
%\centering
%\includegraphics[scale=0.3]{pics//PSD_outdoor_EoA.eps}\label{fig:PSD_outdoor_EoA}
%%\vspace{-5pt}
%\end{minipage}
%}
%\subfigure[AoD PSD.]{
%\begin{minipage}{0.45\linewidth}
%\centering
%\includegraphics[scale=0.3]{pics//PSD_outdoor_AoD.eps}\label{fig:PSD_outdoor_AoD}
%%\vspace{-5pt}
%\end{minipage}
%}
%\subfigure[EoD PSD.]{
%\begin{minipage}{0.45\linewidth}
%\centering
%\includegraphics[scale=0.3]{pics//PSD_outdoor_EoD.eps}\label{fig:PSD_outdoor_EoD}
%%\vspace{-5pt}
%\end{minipage}
%}
%\centering
%%\vspace{-10pt}
%\caption{SAGE estimated angular PSDs at $5.8$ GHz.}\label{fig:AngularPSDoutdoor}
%%\vspace{-10pt}
%\end{figure}

\begin{figure}[htbp]
\centering
%\vspace{-15pt}
\subfigure[AoA AS.]{
\begin{minipage}{0.45\linewidth}
\centering
\includegraphics[scale=0.31]{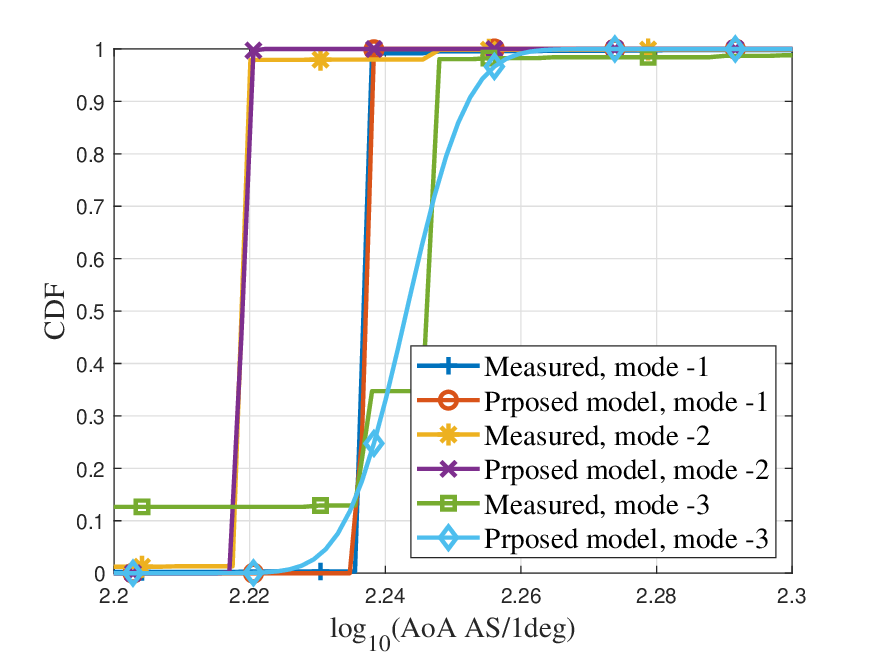}\label{fig:CDF_AS_AoA}
%\vspace{-5pt}
\end{minipage}
}
\subfigure[EoA AS.]{
\begin{minipage}{0.45\linewidth}
\centering
\includegraphics[scale=0.31]{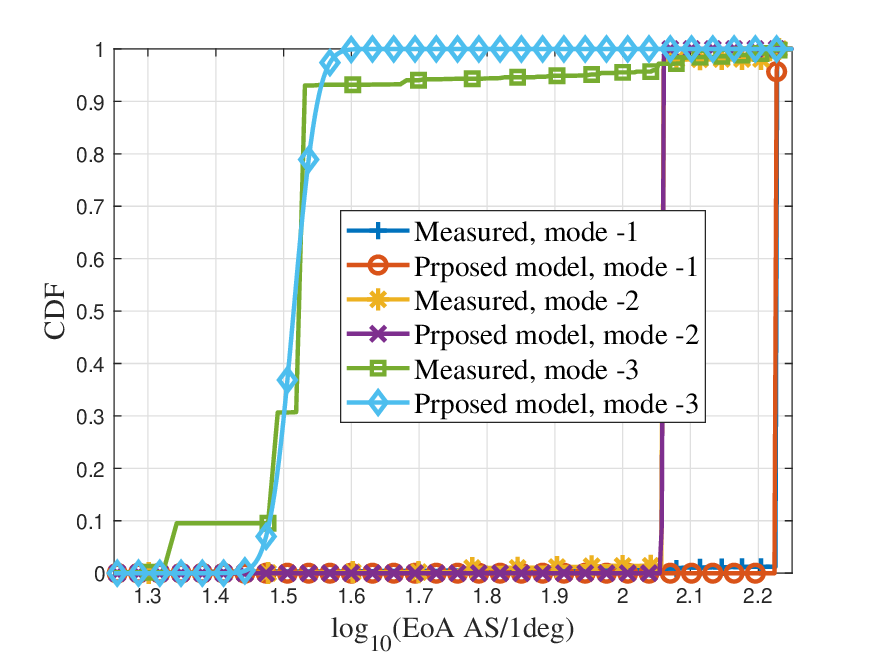}\label{fig:CDF_AS_EoA}
%\vspace{-5pt}
\end{minipage}
}\\
\subfigure[AoD AS.]{
\begin{minipage}{0.45\linewidth}
\centering
\includegraphics[scale=0.31]{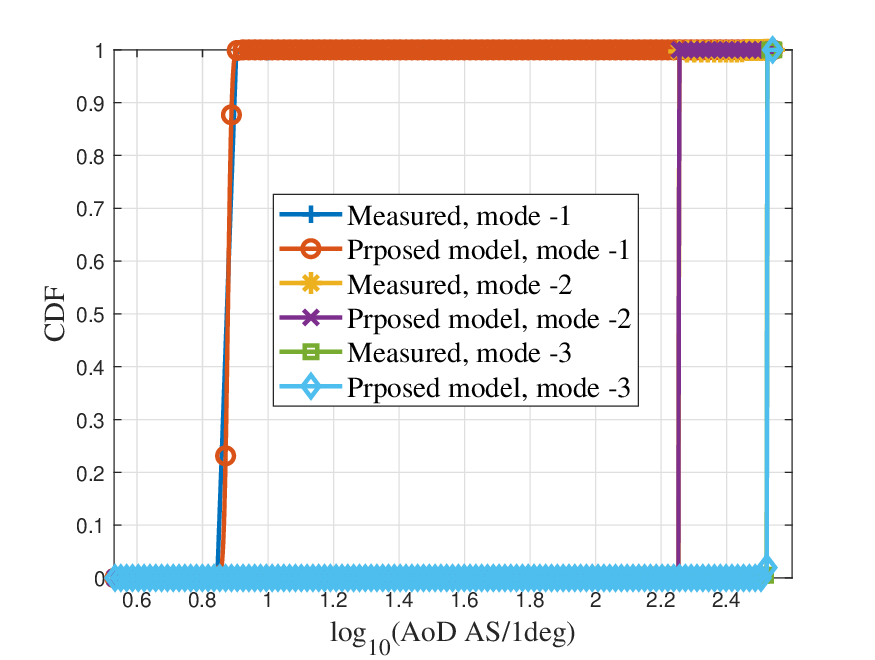}\label{fig:CDF_AS_AoD}
%\vspace{-5pt}
\end{minipage}
}
\subfigure[EoD AS.]{
\begin{minipage}{0.45\linewidth}
\centering
\includegraphics[scale=0.31]{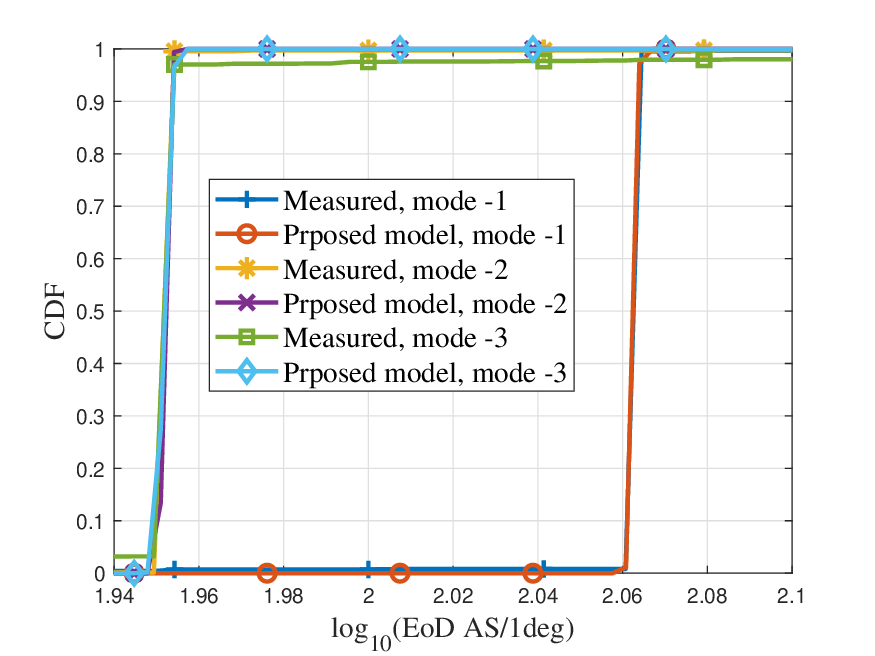}\label{fig:CDF_AS_EoD}
%\vspace{-5pt}
\end{minipage}
}
\centering
%\vspace{-10pt}
\caption{AS CDF at $5.8$ GHz in outdoor LOS scenario.}\label{fig:CDF_AS}
%\vspace{-10pt}
\end{figure}
Figure~\ref{fig:CDF_AS} gives the measured and simulated CDFs of $\log_{10}(\text{AS}/1\ \text{degree})$ for OAM modes $-3$ to $-1$ in an outdoor LOS scenario at $5.8$ GHz with a transmitter-to-receiver distance of $12$ meters. Figure~\ref{fig:CDF_AS} demonstrates that the proposed general channel model is consistent with the measurement data, verifying its accuracy. Additionally, Fig.~\ref{fig:CDF_AS} reveals that the mean values of angular PSDs for OAM modes $-3$ to $-1$ are quite different. However, the variances are all relatively small, indicating that while the central tendencies of the angular spreads differ, their dispersions around the mean are limited.

%\subsection{RMS AS}
\subsection{Channel Correlation Matrix}
\begin{figure}[htbp]
\centering
%\vspace{-15pt}
\subfigure[Indoor LOS scenario at 5.8 GHz with a distance of $9.6$ m.]{
\begin{minipage}{1\linewidth}
\centering
\includegraphics[scale=0.4]{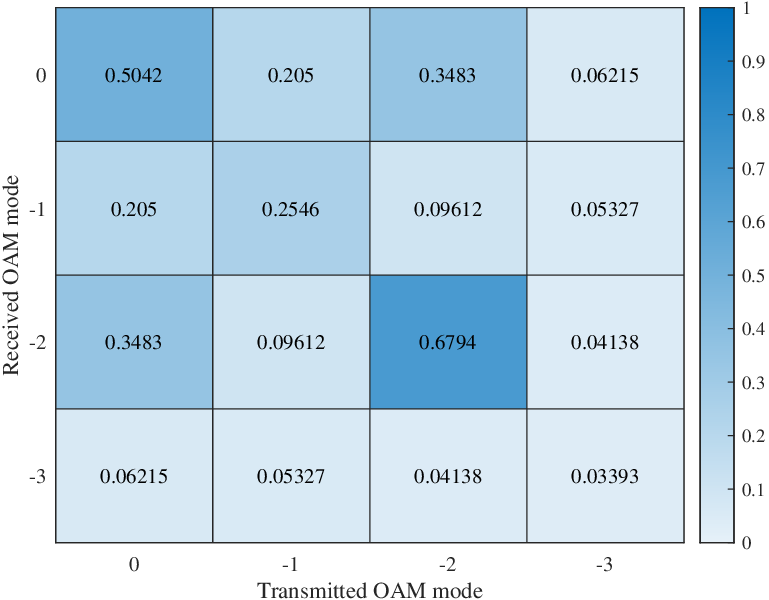}\label{fig:channelCorr5.8G_indoorLOS}
\vspace{5pt}
\end{minipage}
}
\subfigure[Through wall scenario at 5.8 GHz with a distance of $1.15$ m.]{
\begin{minipage}{1\linewidth}
\centering
\includegraphics[scale=0.4]{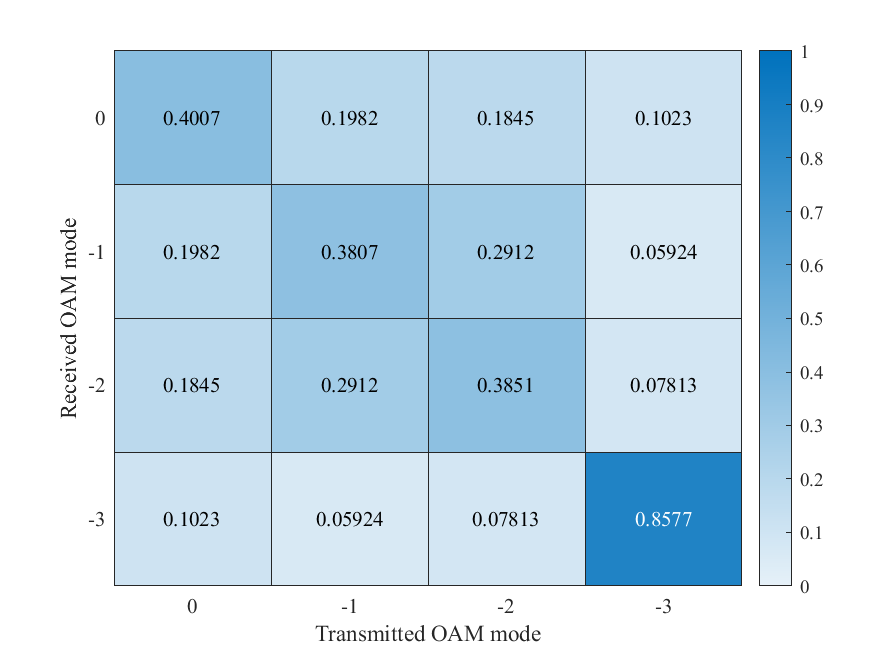}\label{fig:channelCorr5.8G_throWall}
\vspace{5pt}
\end{minipage}
}
\subfigure[Outdoor LOS scenario at 5.8 GHz with a distance of $12$ m.]{
\begin{minipage}{1\linewidth}
\centering
\includegraphics[scale=0.4]{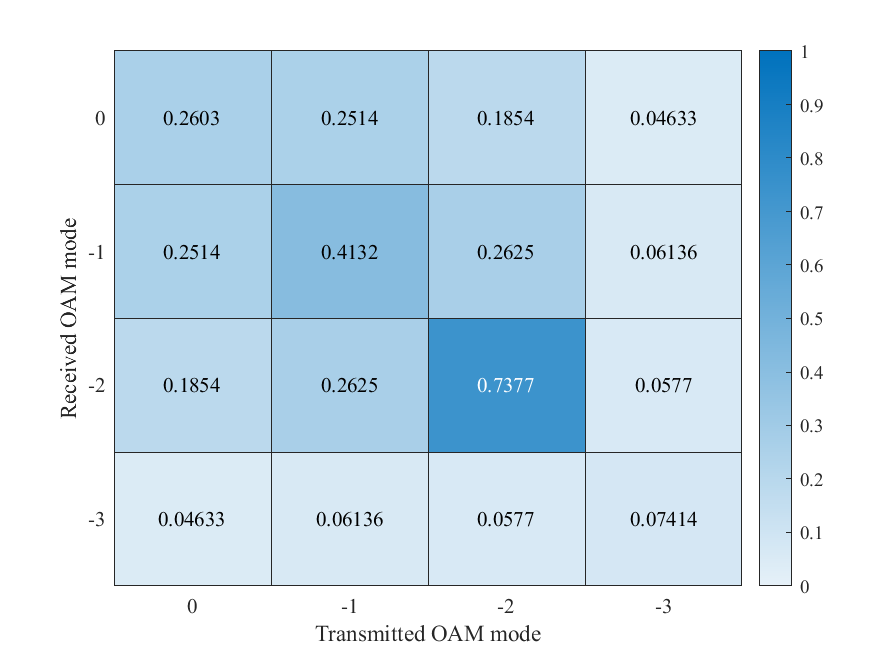}\label{fig:channelCorr5.8G_outdoorLOS}
\vspace{5pt}
\end{minipage}
}
\centering
%\vspace{-10pt}
\caption{Correlation matrices of measured CTFs.}\label{fig:channelCorr}
%\vspace{-10pt}
\end{figure}
\begin{figure}[htbp]
\centering
%\vspace{-15pt}
\subfigure[Indoor LOS scenario at 28 GHz with a distance of $10$ m.]{
\begin{minipage}{1\linewidth}
\centering
\includegraphics[scale=0.42]{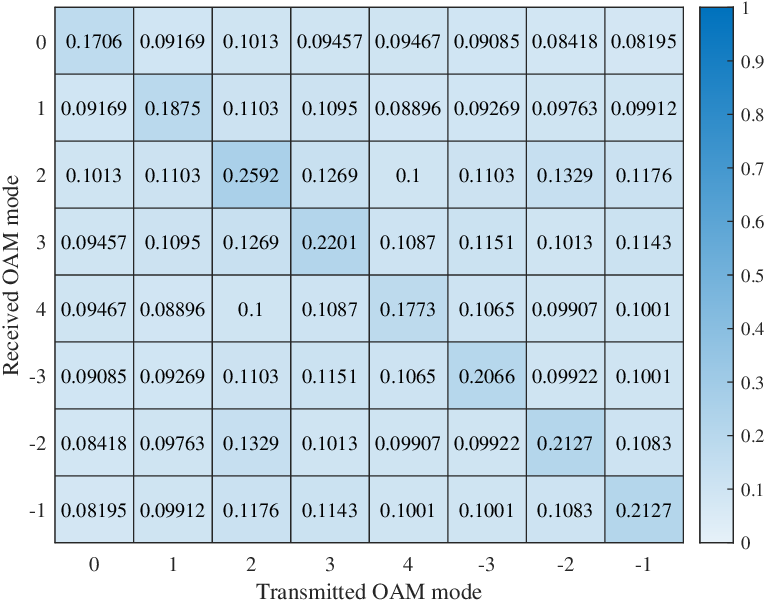}\label{fig:channelCorr28G_indoorLOS}
\vspace{5pt}
\end{minipage}
}
\subfigure[Through wall scenario at 28 GHz with a distance of $3.65$ m.]{
\begin{minipage}{1\linewidth}
\centering
\includegraphics[scale=0.42]{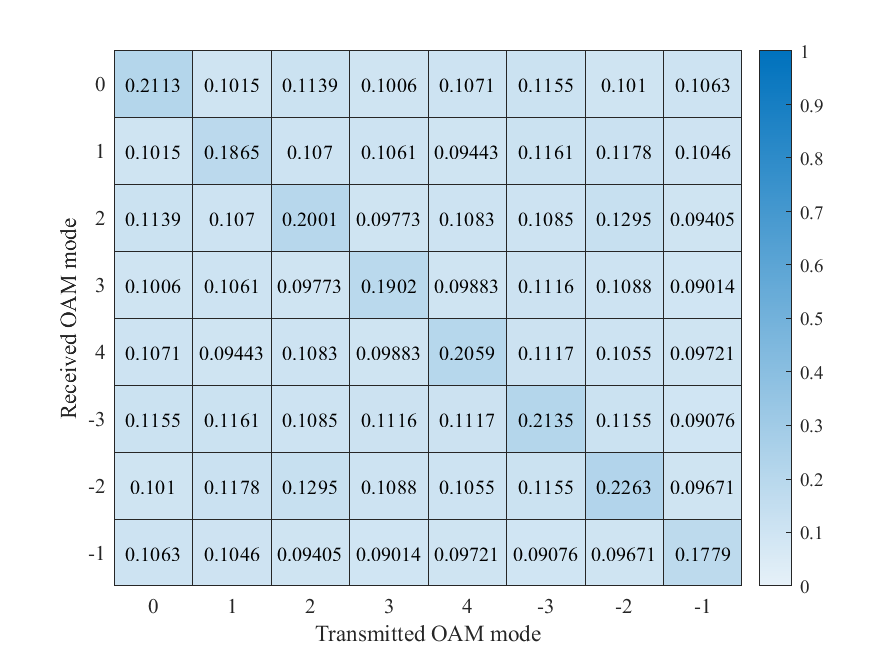}\label{fig:channelCorr28G_throWall}
\vspace{5pt}
\end{minipage}
}
\centering
%\vspace{-10pt}
\caption{Correlation matrices of measured CTFs at $28$ GHz.}\label{fig:channelCorr28G}
%\vspace{-10pt}
\end{figure}
%Figure~\ref{fig:channelCorr} gives the normalized correlation matrices of different scenarios using measured CTFs at $5.8$ and $28$ GHz. The crosstalk between different OAM modes is also evident as shown in Fig.~\ref{fig:channelCorr}. Additionally, the most correlated OAM mode varies across different scenarios and distances at $5.8$ GHz as shown in Figs.\ref{fig:channelCorr5.8G_indoorLOS} to \ref{fig:channelCorr5.8G_outdoorLOS}. We can also find in Figs.\ref{fig:channelCorr28G_indoorLOS} and \ref{fig:channelCorr28G_throWall} that when the transmitting and receiving modes are the same, the correlation is still significantly higher than in other cases.

Figures~\ref{fig:channelCorr} and \ref{fig:channelCorr28G} presents the normalized correlation matrices for various scenarios, using measured CTFs at $5.8$ GHz and $28$ GHz. The crosstalk between different OAM modes is evident in these figure. For the $5.8$ GHz band, Fig.~\ref{fig:channelCorr} demonstrates that the correlation matrices reveal that the most correlated OAM mode varies across different scenarios and distances at $5.8$ GHz. Figure~\ref{fig:channelCorr} shows that environmental factors and the relative positioning of the transmitter and receiver can significantly influence which OAM modes exhibit the highest correlation. For the $28$ GHz band, Fig.~\ref{fig:channelCorr28G} demonstrates that when the transmitting and receiving modes are the same, the correlation is significantly higher compared to other cases. This higher correlation suggests that maintaining identical OAM modes for transmission and reception can enhance signal integrity, even in environments with potential obstacles, such as walls.

%The analysis of these correlation matrices provides valuable insights into the behavior of OAM modes in different frequency bands and scenarios. It highlights the importance of considering environmental factors and the specific characteristics of different frequency bands when designing communication systems that utilize OAM modes.

\subsection{Channel Capacity}
\begin{figure}[htbp]
\centering
%\vspace{-10pt}
\includegraphics[scale=0.45]{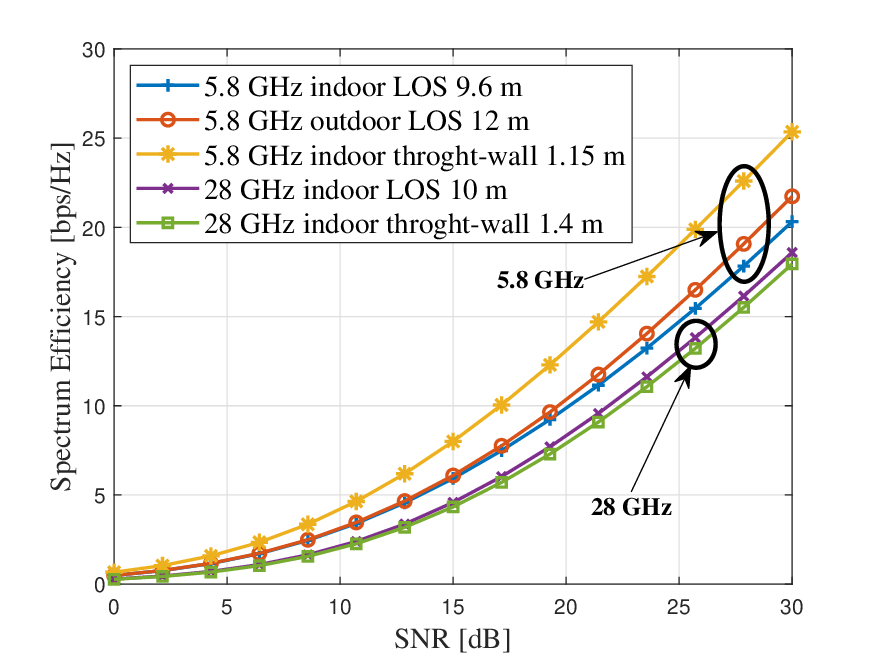}
%\vspace{-15pt}
\caption{Channel capacities of measured CTFs.}\label{fig:capacity}
%\vspace{-5pt}
\end{figure}
%Figure~\ref{fig:capacity} presents the normalized channel capacities for measured CTFs of indoor LOS (9.6 m), indoor through-wall (1.15 m), and outdoor LOS (12 m) scenarios at $5.8$ GHz, as well as indoor LOS (10 m) and indoor through-wall (1.4 m) scenarios at $28$ GHz. Figure~\ref{fig:capacity} shows that although there are twice as many OAM modes used at $28$ GHz compared to $5.8$ GHz, the normalized capacities at $28$ GHz are still lower than those at $5.8$ GHz for the corresponding scenarios. It can also be found in Fig.~\ref{fig:capacity} that the wall attenuates the signal more at $28$ GHz than at $5.8$ GHz by comparing indoor through-wall with indoor LOS scenarios. Additionally, at $5.8$ GHz, the capacity of the outdoor LOS scenario is higher than that of the indoor LOS scenario, which can be attributed to the negative effects of indoor multipaths on OAM transmission.

Figure~\ref{fig:capacity} presents the normalized channel capacities for measured CTFs in various scenarios. Specifically, it includes data for indoor LOS ($9.6$ meters), indoor through-wall ($1.15$ meters), and outdoor LOS ($12$ meters) scenarios at $5.8$ GHz, as well as indoor LOS ($10$ meters) and indoor through-wall ($1.4$ meters) scenarios at $28$ GHz. Despite the use of twice as many OAM modes at $28$ GHz compared to $5.8$ GHz, Fig.~\ref{fig:capacity} shows that the normalized capacities at $28$ GHz are still lower than those at $5.8$ GHz for the corresponding scenarios. This observation can be attributed to the higher frequency's increased susceptibility to signal attenuation and scattering, leading to a reduction in channel capacity. Moreover, the figure highlights that walls attenuate the signal more at $28$ GHz than at $5.8$ GHz. This is evident when comparing the indoor through-wall and indoor LOS scenarios, where the normalized capacities drop more significantly at the higher frequency. This pronounced attenuation at $28$ GHz is likely due to the higher frequency signals' greater sensitivity to obstructions. Additionally, at $5.8$ GHz, the capacity of the outdoor LOS scenario is higher than that of the indoor LOS scenario. This can be attributed to the negative effects of indoor multipaths on OAM transmission. In indoor environments, reflections and scattering from walls and other objects can cause signal degradation and interfere with the transmission of OAM modes, reducing overall channel capacity.

%These findings emphasize the challenges and considerations necessary for optimizing OAM-based communication systems, particularly at higher frequencies. Understanding the impact of environmental factors and frequency-specific behavior on channel capacity is crucial for designing efficient and robust communication systems.

\section{Conclusion}\label{sec:conclustion}
In this paper, we measured and modeled OAM channels in the $5.8$ GHz and $28$ GHz bands. We first introduced the systems and measurement scenarios. Then, we proposed a 3D-GBSM for the OAM channel, modified from the twin-cluster-based 3D-GBSM, to describe various stochastic properties of OAM channels. Based on our proposed OAM channel model, we introduced the measurement data processing procedure. We then derived the MPC parameters of the OAM 3D-GBSM according to the processed data in the given measurement environment. We also calculated the second-order stochastic properties and the average path losses. Finally, we provided the simulation channel generation scheme for the given measurement environment. Additionally, we gave some channel measurement and simulation results for verifying our channel model as well as providing some further insights.

%\begin{appendices}
%\section{Proof for Theorem~\ref{the:channel_matrix}}\label{pro:multi-coil channel}
%
%\end{appendices}

\bibliographystyle{IEEEtran}
\bibliography{References}

\ifCLASSOPTIONcaptionsoff
  \newpage
\fi

\end{document}